\documentclass{elsart}
\usepackage{natbib}
\usepackage{graphicx}
\usepackage{amssymb}

\begin{document}

\begin{frontmatter}

\title{Clusters in simple fluids} 

\author{N. Sator\thanksref{ad}} 
\ead{sator$@$na.infn.it}

\thanks[ad]{Present address: Laboratoire de Physique Th\'eorique des
Liquides, Universit\'e Pierre et Marie Curie, 4, Place Jussieu 75252
Paris Cedex 05 France.}

\address{Department of Physics and Astronomy, McMaster University,
Hamilton, Ontario L8S 4M1, Canada,\\ Dipartimento di Scienze Fisiche,
INFM Napoli, Universit\`a di Napoli ``Federico II'', Complesso
Universitario di Monte Sant'Angelo, Via Cintia, I-80126 Napoli, Italy}

\begin{abstract}

\noindent
This article concerns the correspondence between thermodynamics and
the morphology of simple fluids in terms of clusters. Definitions of
clusters providing a geometric interpretation of the liquid-gas phase
transition are reviewed with an eye to establishing their physical
relevance. The author emphasizes their main features and basic
hypotheses, and shows how these definitions lead to a recent approach
based on self-bound clusters. Although theoretical, this tutorial
review is also addressed to readers interested in experimental aspects
of clustering in simple fluids.
\end{abstract}



\end{frontmatter}

\tableofcontents

\section{Introduction}
\label{sec:introduc}

Simple fluids are classical systems in which chemically inert
particles interact through a pairwise potential. The main features of
the potential are the hard-core repulsion and the short-ranged
attraction which give rise to thermal phase transitions. For example,
noble gases and alkali metals are often modeled as simple fluids.

Moreover, attraction between particles promotes cluster formation. It
is then sensible to look for a relation between thermodynamics and the
morphology of a fluid in terms of clusters. In particular, a geometric
interpretation of the liquid-gas phase transition is an important
issue in statistical mechanics. To put it more precisely, there are
two main issues: firstly to understand condensation as the sudden
formation of a macroscopic cluster, and secondly to describe the
morphology of the fluid at the critical point with a view to infer its
thermodynamic properties.

To begin, it must be said that a geometric description of a fluid is
outside the scope of thermodynamics. In addition to standard tools of
thermal statistical physics, one has to define what is a cluster. At
first sight, it may seem obvious to define a cluster as a set of
particles close to each other and far from other clusters. However,
this definition becomes quite ambiguous at high density, even below
the critical density. Therefore, more sophisticated definitions of
clusters were introduced, giving the geometric description of simple
fluids a long and rich story.

This article presents a comprehensive review of proposed definitions
of clusters, with a view to establish a relation between
thermodynamics and a geometric description of the fluid. However, the
reader will see that this purpose will take us beyond this particular
topic. On the one hand, our aim is to outline the main features of
these various clusters. In particular, we highlight the hypotheses on
which their definitions are built, in order to discuss their physical
relevance. Indeed, some of these definitions were designed as
mathematical tools to study thermodynamic properties. However, these
clusters are often used in literature as though one could observe them
experimentally. On the other hand, we clarify the relations between
these definitions by following a chronological order so as to avoid
frequent misunderstandings. Hence, we develop a new point of view
about the Kert\'esz line associated with the Coniglio-Klein
clusters. From this standpoint, we present a recent approach in terms
of self-bound clusters that provides a physical interpretation of the
liquid-gas phase transition and suggests the existence of a
percolation line in the supercritical phase of simple fluids.

This article is aimed not only at readers interested in fundamental
aspects of the relation between thermodynamics and geometry, but also
at experimentalists who deal with clusters in various fields of
physics. In particular, we think of physicists who search for vestiges
of phase transitions in small systems like atomic nuclei, and
aggregates.

The plan of the review is as follows. In section \ref{sec:thermo}, we
recall the definition of a simple fluid and describe its typical phase
diagram. Section \ref{sec:liquidgas} is devoted to definitions of
clusters based on the ``perfect gas of clusters model'' developed in
appendix \ref{sec:pgc}.  Section \ref{sec:lattice} discusses
microscopic definitions of clusters proposed in the framework of the
lattice-gas model by means of percolation theory. In section
\ref{sec:self}, we are interested in self-bound clusters defined by
energetic criteria. We present the correspondence between the phase
diagram and their cluster size distribution, in the framework of the
lattice-gas model, and by using numerical simulations of a
Lennard-Jones fluid. Finally, section \ref{sec:conc} contains a
summary and we close by discussing several open questions.

\section{From thermodynamics to the morphology of simple fluids}
\label{sec:thermo}

A simple fluid\footnote{For a comprehensive review about simple
liquids, we refer the reader to the book of Hansen and McDonald
\citeyearpar{Hans86}.} is a classical system composed of $N$
structureless particles of mass $m$, interacting through a two-body
additive central potential $u(r_{ij})$, where $r_{ij}$ is the distance
between particles $i$ and $j$. The potential must have a repulsive
hard-core and a short range attraction that vanishes faster than
$-1/r_{ij}^3$ to ensure the thermodynamic limit to exist in three
dimensions \citep{Ruel69}. The Lennard-Jones potential, for instance,
satisfies these conditions:
\begin{equation} \label{lj} u(r_{ij})=4\epsilon
[(\frac{\sigma}{r_{ij}})^{12}-(\frac{\sigma}{r_{ij}})^{6}]
\end{equation}
where the two constants $\epsilon$ and $\sigma$ fix the energy and
length scales respectively. The Hamiltonian of the fluid is the sum of
a kinetic and an interaction energy term:
\begin{equation} \label{ea} \mathcal{H}(\{\vec r_{i}\},\{\vec p_{i}\})=
\sum_{i=1}^N\frac{{\vec{p_{i}}}^2}{2m}+\sum_{i<j}u(r_{ij})
\end{equation} 
where $\vec r_{i}$ and $\vec{p_{i}}$ are the position and momentum
coordinates of the $i^{th}$ particle, and $r_{ij}=|\vec r_{i}-\vec
r_{j}|$. In the canonical ensemble, for a given volume $V$ and
temperature $T$, the partition function of the fluid is given by
$$
Q_{N}(T,V)=\frac{1}{N!h^{3N}}\int e^{-\beta \mathcal{H}(\{\vec r_{i}\},\{\vec
p_{i}\})} d\vec r_{1}\dots d\vec r_{N}d\vec p_{1} \dots d\vec
p_{N} 
$$
where $\beta=1/kT$, $k$ is the Boltzmann's constant and $h$ the Planck's
constant. Integrations over the momentum coordinates are
straightforward and allow us to write the partition function as
\begin{equation} Q_{N}(T,V)=\frac{1}{N!\lambda^{3N}}Z_{N}(T,V) \label{e3} 
\end{equation}
where \begin{equation} \label{long} \lambda=\frac{h}{\sqrt{2\pi mkT}}
\end{equation} and
\begin{equation} \label{zn} Z_{N}(T,V)= \int_{V} 
\prod_{i<j}e^{-\beta u(r_{ij})} d\vec r_{1} \dots d\vec r_{N}.
\end{equation} 
Equation (\ref{long}) defines the thermal wavelength. The
thermodynamics of the fluid is then determined by the behaviour of the
configuration integral $Z_{N}(T,V)$ in the thermodynamic limit.

\begin{figure}[h!]
\begin{center}
\includegraphics[angle=0,scale=.45]{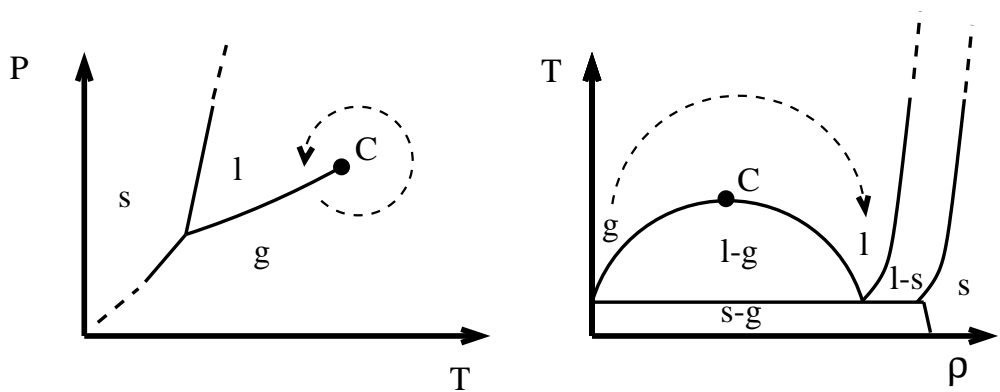}
\end{center}
\caption{\it Schematic phase diagram of a simple fluid in the $P-T$
plan (left) and the $T-\rho$ plan (right), where $P$, $T$ and $\rho$
are respectively the pressure, temperature and density of the
system. The solid, liquid and gas phases are respectively noted $s$,
$l$, and $g$. The critical point is C. The dashed line with the arrow
represents a path in the phase diagram that does not lead to
singularities of the partition function.}
\label{fig1}
\end{figure}

The typical phase diagram of the system is shown in
Fig. \ref{fig1}. In the $P-T$ diagram, coexistence curves separate the
plane into three regions corresponding to the solid, liquid, and gas
states. The liquid-gas coexistence curve stops at the critical point
($C$). Crossing this curve, the system undergoes a first order phase
transition signaled by a latent heat and the coexistence of a low
density phase, the gas, and a high density phase, the liquid. In other
words, compressing a gas at constant temperature $T$ below the
critical temperature $T_{c}$, a first drop of liquid appears when, in
the phase diagram, one reaches the liquid-gas coexistence curve at the
pressure $P_{cond}(T)$, at the corresponding density
$\rho_{cond}(T)$. This phenomenon is called ``condensation''. On the
other hand, just at the critical point, a continuous phase transition
occurs, while the compressibility and the specific heat diverge. The
critical behaviour is known to be in the universality class of the
Ising model (see for instance \citealp{Gold92}). In the $T-\rho$
diagram, above the bell-shaped liquid-gas coexistence curve, there is
one single fluid phase called the ``supercritical
phase''.\footnote{The supercritical phase of a fluid (water) was
discovered by the Baron Cagniard de La Tour \citeyearpar{Cagn22}.}
Because the thermodynamic properties of the system vary smoothly along
any path that does not cross a coexistence curve, it is possible to
pass continuously from the gas to the liquid phase by following a path
such as the dashed line shown in Fig. \ref{fig1}.

In the gas, liquid, and supercritical phases, the system is
homogeneous, density is uniform on average. However, attractive
interaction between particles contributes to forming localized
morphological structures, the clusters. It must be emphasized that in
order to study the morphology of a fluid, we need to give a definition
of clusters in addition to standard thermodynamic tools. Indeed, the
partition function depends only on the sum of the two-particle
interactions, without discerning if a particle belongs to a cluster or
not. In other words, one could arbitrary distribute the $N$ particles
among a set of clusters, while the partition function and all the
thermodynamic properties would not be altered. Of course, the choice
of the definition of clusters has no influence on the thermodynamic
behaviour of the fluid, but depends on the problem we want to study.

Historically, the first motivation for studying the morphology of a
fluid was to interpret condensation as the sudden formation of a
macroscopic cluster. At that time, the purpose was especially to
determine thermodynamic quantities with the help of ``effective''
clusters used as a convenient and powerful tool. These first
definitions of clusters are based on the ``perfect gas of clusters
model''.

\section{Perfect gas of clusters}
\label{sec:liquidgas}

The basic idea is to assume that an imperfect fluid made up of
interacting particles can be considered as a perfect gas of clusters
at thermodynamic and chemical equilibrium: clusters do not interact
with each other and do not have any volume. We shall discuss the
validity of the perfect gas of clusters approximation at the
microscopic level when we tackle the problem of self-bound clusters in
section \ref{these}.

Let us suppose we deal with a collection of non-interacting clusters,
without knowing how we can build them as sets of interacting
particles. Clusters of given size $s$ are characterized by their mass
$ms$, a chemical potential $\mu_s$, and a partition function
$q_{s}(T,V)$. We show in appendix \ref{sec:pgc} that the pressure, the
density, and the cluster size distribution, that is the mean number
of clusters of size $s$, are respectively given by
\begin{eqnarray}
&\beta P& = \frac{1}{V} \sum_{s=1}^{\infty} q_{s}z^{s}
\label{fbp} \\ &\rho& = \frac{1}{V}\sum_{s=1}^{\infty} s q_{s}z^{s}
\label{fbr} \\ &n_{s}(T,z)& = q_{s}z^s \label{nsf}
\end{eqnarray}
where $z=e^{\beta \mu}$ is the fugacity and $\mu$ the particle
chemical potential. The equation of state of the fluid can be obtained
by eliminating $z$ in the pressure expression with the help of
Eq. (\ref{fbr}). It is interesting to write Eq. (\ref{fbp}) as a
function of the cluster size distribution given by Eq. (\ref{nsf}):
\begin{equation}
\beta P V = \sum_{s=1}^{\infty} n_{s}.
\label{egp}
\end{equation}
Equation (\ref{egp}) shows strikingly that the imperfect fluid can be
seen as a perfect gas composed of $m_0=\sum_{s=1}^{\infty} n_{s}$
non-interacting clusters.

According to Eqs. (\ref{fbp})-(\ref{nsf}), thermodynamic quantities,
as well as the cluster size distribution, are completely determined by
the partition function of the clusters. Now, two points of view can be
considered. In the pioneer theory of condensation proposed by Mayer
and his collaborators in 1937, clusters result from an exact
enumeration of all the possible combinations of interacting particles
(sec. \ref{sec:mayer}). It is equivalent to giving a formal expression
of $q_s(V,T)$ that is impracticable to handle for large size
clusters. On the other hand, Frenkel, Band (sec. \ref{sec:frenkel}),
and Fisher (sec. \ref{sec:fisher}) propose a phenomenological
expression of the partition function $q_s(V,T)$, and assume that
clusters do not interact whatever the density.

\subsection{Mayer clusters}
\label{sec:mayer}

Mayer's theory of condensation is described in many textbooks of
statistical physics (see for example \citealp{Pathria72,Huang87}) and
in Mayer's articles
\citep{May37,May-Ack37,May-Har38,Har-May38,May-May40}. Here, we just
want to show how Mayer clusters can be seen as a particular case of
the perfect gas of clusters model.

Originally, this theory was intended to express the exact equation of
state of a real fluid, made up of interacting particles, as a series
expansion in the density. To this end, Mayer decomposes the partition
function (\ref{e3}) into the sum over all the possible partitions of
particles into independent mathematical clusters. The first step is to
introduce a function of the potential that is significant only if
particles are close to each other:
$$
f_{ij}=e^{-\beta u(r_{ij})}-1.
$$
The function $|f_{ij}|$ is bounded everywhere, unlike the potential
$u(r_{ij})$, and tends to $0$ when $r_{ij}$ becomes large compared to
the interaction range. Using this function as a small parameter, one
can write the configuration integral (\ref{zn}) as a diagrammatic
expansion by associating an $N$-particle graph with each
term.\footnote{According to Huang \citeyearpar{Huang87}, this is the
first graphical representation of a perturbation series expansion in
physics.} A graph consists of vertices and bonds: to each particle
corresponds a vertex, and the function $f_{ij}$ is represented by a
bond connecting vertices $i$ and $j$. A set of pair by pair connecting
vertices is called a ``Mayer cluster''. Each graph, or term of the
expansion, is characterized by a cluster size distribution.

By collecting the terms having the same cluster size distribution, one
obtain a new expression of the partition function that can be
simplified by moving over to the grand canonical ensemble. In the
thermodynamic limit, pressure and density are eventually written as
series expansion in the fugacity. As a result, calculations lead to
the expressions (\ref{fbp}) and (\ref{fbr}), in which $q_s/V$ is
replaced by $\widetilde{b}_{s}(T)/\lambda^3$, where coefficients
$\widetilde{b}_{s}(T)$ are the so-called ``cluster integrals''
introduced by Mayer, and $\lambda$ is the thermal wavelength of the
particle defined by Eq. (\ref{long}).

In order to locate the condensation point, that is the density at
which a macroscopic cluster appears, we have to examine the cluster
size distribution given by Eq. (\ref{nsf}). At a given temperature,
the behaviour of the coefficients $\widetilde{b}_{s}(T)$ for $s>>1$
shows that with increasing density, $n_{s}(T,\rho)$ becomes non-zero
when $\rho>\rho_{cond}$. A macroscopic cluster appears, and Mayer
identifies this particular density $\rho_{cond}$ with the density of
condensation. However, Yang and Lee \citeyearpar{Yan-Lee52} proved
that Mayer's theory is exact in the gas phase, for $\rho<\rho_{cond}$,
but cannot be carried forward into the liquid phase.
 
By definition, Mayer clusters form a perfect gas of clusters and are
designed to evaluate thermodynamic quantities. Yet, it can be shown
that $\widetilde{b}_{s}(T)$, and therefore the cluster size
distribution, may be negative in certain thermodynamic
conditions. There then appears to be no physical interpretation of
these clusters \citep{Fish67a,Pathria72}.

\subsection{Frenkel clusters}
\label{sec:frenkel}

In contrast to Mayer's theory, the basic idea of Frenkel's
model\footnote{A similar model has been independently proposed by Bijl
\citeyearpar{Bijl38} and Band \citeyearpar{Band39a}.}
\citep{Frenk39a,Frenk39b} is to use directly the perfect gas of
cluster model by choosing a phenomenological expression of the
partition function of clusters. Afterward, this model was modified to
allow for the volume of the clusters \citep{Band39b,Stil63}, but
attraction between clusters was still overlooked.

By disregarding the degrees of freedom associated with the particles,
Frenkel assumes the clusters of size $s$ to be compact and writes
their potential energy $Ep_s$, for $s$ sufficiently large, as the sum
of a bulk and a surface term
$$
Ep_{s}(T)=-e_{v}s+e_{a} s^{\frac{2}{3}} \qquad \textrm{for} \qquad
s>>1
$$
where $e_{v}>0$ and $e_{a}>0$ are the bulk and surface potential
energy by particle respectively. Frenkel does not take into account
the entropy of the clusters and infers the partition function by
integrating over the position and momentum coordinates of the center
of mass of the cluster:
$$
q_{s}(T,V) =V {(\frac{\sqrt{s}}{\lambda})}^3e^{-\beta Ep_{s}} =
\frac{V}{\lambda^3}s^{\frac{3}{2}}e^{\beta
(e_{v}s-e_{a}s^{\frac{2}{3}})}
$$
where $\lambda/\sqrt{s}$ is the thermal wavelength of a cluster of
size $s$ (see Eq. (\ref{long})). According to Eq. (\ref{nsf}), the
Frenkel cluster size distribution is given by
\begin{equation}
\frac{n_{s}(T,z)}{V}=\frac{s^{\frac{3}{2}}}{\lambda^3}y^{s}x^{s^{\frac{2}{3}}}
\label{fbns}
\end{equation} with
$$
y(T,z)=ze^{\beta e_{v}} \qquad \textrm{and} \qquad x(T)=e^{-\beta
e_{a}}.
$$
The function $y$ depends on the temperature and the density through
$z$. The function $x$ is independent of the density and always less
than 1. For a given temperature $T$, $x$ is fixed, the cluster size
distribution depends on density only through $y(T,z)$. When $y<1$,
$n_{s}$ decreases exponentially for $s>>1$: there is no macroscopic
cluster, this corresponds to the gas phase. For $y>1$, $n_{s}$
decreases as long as $s$ is smaller than a particular size, and then
increases exponentially: a macroscopic cluster appears, which
indicates the formation of the liquid phase.  The density at the
condensation point is then $\rho=\rho_{cond}(T)$ such that
$y=y_{cond}=1$.

The results of Frenkel's phenomenological model are for the most part
the same as those of the Mayer's theory of condensation. The main
advantage is that Frenkel clusters do not have the pathology of the
Mayer clusters: the cluster size distribution (\ref{fbns}) is positive
whatever the temperature and density. However, this model does not
allow one to locate the critical point and describe the morphology of
the fluid in this state.

\subsection{Fisher droplets}
\label{sec:fisher}

As Frenkel's model, Fisher's droplet model
\citep{Fish67a,Fish67b,Fish71} is based on the perfect gas of clusters
approximation. However, Fisher writes the partition function of the
clusters of size $s$ with the help of additional features: he allows
for the entropy of the clusters, clusters are not assumed to be
compact, and a corrective term varying like $\ln s$ is added to the
free energy of the clusters of size $s$, in addition to the surface
and volume terms.

Here, we shall not follow Fisher's original calculations
\citep{Fish67a}. We rather propose a simplified and more direct way of
calculating the partition function of clusters of size $s$
\citep{Fish71}.

Let us first write the free energy of the clusters of size $s$. The
mean internal energy and entropy of a cluster of size $s>>1$, with a
mean surface area $A_s$, are written as the sum of a surface and
volume term:
\begin{eqnarray*}
U_{s} &=&-u_{v}s+u_{a} A_s \\ 
S_{s} &=&s_{v}s+s_{a} A_s
\end{eqnarray*}
where $u_{v}>0$ and $s_{v}>0$ are the volume energy and entropy by
particle respectively, and $u_a > 0$ and $s_a>0$, the surface energy
and entropy by particle. Fisher does not assume the clusters to be
compact and introduces a parameter $\sigma$ to characterize the mean
surface of the clusters of size $s$:
$$
A_s(T)=a_{0}(T)s^{\sigma} \qquad \textrm{with} \qquad 0<\sigma<1.
$$
The surface of a three-dimensional cluster should lie between the
surface of a compact object ($\sigma=2/3$) and the surface of
a chain ($\sigma=1$). Values of $\sigma$ less than $2/3$ are
interpreted by Fisher as an effective evaluation of the interaction
between clusters.

Fisher also adds to the free energy a corrective logarithmic term
which has no obvious physical interpretation\footnote{This term is
supposed to allow for the surface undulation of the clusters
\citep{Essam63,Fish71}.} \citep{Bind76b}. The weight of this term is
given by a second parameter $\tau$. We shall see later that this term
plays a crucial role in this model. The free energy $F_{s}(T,V)$ of a
cluster of size $s$ is then given by
$$
-\beta F_{s}(T,V)=\beta(u_v+s_v T)s-\beta a_{0}(u_a-s_a
 T)s^{\sigma}-\tau\ln s +\ln c_{0}V.
$$
The term proportional to $\ln V$ results from the integration over the
position of the center of mass of the cluster and $c_0$ is a
constant. The partition function of a cluster of size $s$ is
eventually given by
$$
q_{s}(T,V) = e^{-\beta F_{s}(T,V)} = c_{0}Vs^{-\tau}e^{\beta(u_v+s_v
T)s-\beta a_{0}(u_a-s_a T)s^{\sigma}}.
$$
On the basis of the ``perfect gas of clusters model'' presented in
appendix \ref{sec:pgc}, the pressure, density, and the cluster size
distribution are written as
\begin{eqnarray}
\beta P &\equiv& \pi(T,z) = \frac{1}{V} \sum_{s=1}^{\infty} q_{s}z^{s} = 
c_{0}\sum_{s=1}^{\infty}y^{s}x^{s^{\sigma}} s^{-\tau} \label{fpkt}\\ 
\rho &=& \frac{1}{V}\sum_{s=1}^{\infty} s q_{s}z^{s}=
c_{0}\sum_{s=1}^{\infty}sy^{s}x^{s^{\sigma}} s^{-\tau} \label{fro}\\
\frac{n_{s}}{V} &=& \frac{q_{s}}{V}z^s =c_{0} y^{s}x^{s^{\sigma}} s^{-\tau}
\label{fns}
\end{eqnarray}
where
\begin{eqnarray*}
y(T,z)& = & z e^{\beta(u_v+s_v T)} \\
x(T) & = & e^{-\beta a_{0}(u_a-s_a T)}.
\end{eqnarray*}
Following Fisher, we have introduced the series $\pi(T,z)$ in
Eq. (\ref{fpkt}) and we define the derivative of order $n$ by
$$
\pi^{(n)}(T,z)=(z\frac{\partial}{\partial
z})^{(n)}\pi(T,z)=c_{0}\sum_{s=1}^{\infty}y^{s}x^{s^{\sigma}}
s^{n-\tau}.
$$
The condensation point is determined by the appearance of a
macroscopic cluster. According to the value of $x$, two cases are
possible: if $x<1$, as in Frenkel's model, the density of the
condensation point $\rho_{cond}$ is given by $y=y_{cond}=1$, that is
$z_{cond}=e^{-\beta(u_v+s_v T)}$. On the other hand, if $x\ge1$, the
cluster size distribution increases when $y>1$ and the series
(\ref{fpkt}) and (\ref{fro}), which give the pressure and density,
diverge. Therefore, condensation only happens when $x<1$, that is for
$T<T_c={u_a}/{s_a}$. This upper limit on the condensation temperature
is interpreted as the critical temperature. Consequently, Fisher's
model is not valid for $T>T_{c}$ or $\rho>\rho_{cond}$ ($x>1$ or
$y>1$).

Despite the basic approximation of the perfect gas of clusters model,
Fisher extends the validity of his model to the case of high densities
and studies the vicinity of the critical point. At the critical point,
$y=x=1$, the critical density is given by
$$
\rho_{c}=\pi^{(1)}(T_{c},z_{cond}) =c_{0}\sum_{s=1}^{\infty} s^{1-\tau}.
$$
The convergence of this series implies $\tau>2$. The behaviour of the
thermodynamic quantities in the neighbourhood of the critical point is
obtained by moving over to the continuous limit. The series
$\pi^{(n)}(T,z)$ can be written for $z=z_{cond}$:
$$
\pi^{(n)}(T,z_{cond})=c_{0} \sum_{s=1}^{\infty} s^{n-\tau}
x^{s^{\sigma}} \sim c_{0} \int_{0}^{\infty} s^{n-\tau} e^{-\theta
s^{\sigma}} ds
$$
where $\theta=\ln x^{-1}=\beta a_{0}s_a(T_{c}-T)$. Performing the
change of variable $t=\theta s^{\sigma}$, we have
$$
\pi^{(n)}(T,z_{cond}) \sim \frac{c_{0}}{\sigma}
\theta^{\frac{\tau-n-1}{\sigma}} \Gamma(\frac{n-\tau+1}{\sigma}) \sim
(T_{c}-T)^{\frac{\tau-n-1}{\sigma}}.
$$
We show in appendix \ref{sec:pgc} how thermodynamic quantities are
linked to the derivatives $\pi^{(n)}$. According to the value of $n$,
we infer the behaviour in the vicinity of the critical point of the
specific heat $C_{v}$ ($n=0$), the density ($n=1$), the
compressibility $\chi$ ($n=2$), and the pressure ($n=0$):

\parbox{10cm} 
{\begin{eqnarray*} C_{v} &\sim& (T_{c}-T)^{-\alpha_{T}}
\qquad \textrm{with} \qquad \alpha_{T}=2-\frac{\tau-1}{\sigma} \\
\rho_{c}-\rho &\sim& (T_{c}-T)^{\beta_{T}} \qquad ~~ \textrm{with}
\qquad \beta_{T}=\frac{\tau-2}{\sigma} \\ \chi &\sim&
(T_{c}-T)^{-\gamma_{T}} \qquad \textrm{with} \qquad
\gamma_{T}=\frac{3-\tau}{\sigma} \\ P_{c}-P &\sim&
(\rho_{c}-\rho)^{\delta_{T}} \qquad ~~~ \textrm{with} \qquad
\delta_{T}=\frac{1}{\tau-2} 
\end{eqnarray*}}\hfill
\parbox{1cm}{\begin{eqnarray}\label{eqscal} \end{eqnarray}} 

\noindent
Fisher's droplet model allows one to write the thermodynamic critical
exponents $\alpha_{T}$, $\beta_{T}$, $\gamma_{T}$, and $\delta_{T}$ as
functions of the two geometric parameters $\sigma$ and
$\tau$. Furthermore, the scaling relations between the thermal
exponents \citep{Fish67b}, which had made this model well-known,
follow easily from Eqs. (\ref{eqscal}):
\begin{eqnarray} 
\alpha_T + 2\beta_T+\gamma_{T}=2 \label{eqscal1} \\
\beta_{T}(\delta_{T}-1)=\gamma_{T} \label{eqscal2}.
\end{eqnarray}
Let us now show the relation between the thermodynamics and the
geometric description of the fluid at the critical point. From
Eq. (\ref{fns}), the cluster size distribution is written for $x=y=1$
$$
\frac{n_{s}}{V}=c_{0} s^{-\tau}.
$$
We now see that the term in $\ln s$, added by Fisher to the free
energy, is essential at the critical point. Without this term, we
cannot infer the scaling relations (\ref{eqscal1})-(\ref{eqscal2}) and
the power law behaviour of the cluster size distribution. It has to be
pointed out that the relation between a power law cluster size
distribution and the critical opalescence, observed at the critical
point, is not clear today. A cautionary remark: this phenomenon, due
to the divergence of the density fluctuations, is explained by
thermodynamics without any reference to a geometric description of the
fluid \citep{Pathria72,Chandler87}.

The two geometric exponents $\tau$ and $\sigma$ can be written as
functions of the thermodynamic critical exponents by reversing
Eqs. (\ref{eqscal}):
\begin{eqnarray*} 
\sigma &=& \frac{1}{\beta_{T} \delta_{T}}=\frac{1}{\beta_{T} +\gamma_{T}} \\
\tau &=& 2+ \frac{1}{\delta_{T}}=2+\frac{\beta_{T}}{\beta_{T}+\gamma_{T}}.
\end{eqnarray*}
In order to respect the critical behaviour of the thermodynamic
quantities, exponents defined by Eqs. (\ref{eqscal}) must be
positive. Consequently, the values of the geometric exponents are
restricted to $2<\tau<3$, and $\sigma>0.5$. Moreover, the inflection
point\footnote{In the vicinity of the critical point, $P_{c}-P \sim
(\rho_{c}-\rho)^{\delta_{T}}$, and the second derivative of the
pressure as a function of the density must vanish, then ${\partial^{2}
P}/{\partial \rho^{2}} \sim (\rho_{c}-\rho)^{\delta_{T}-2}=0$. As a
result, $\delta_{T}>2$.} of the critical isotherm implies
$\delta_{T}>2$. Therefore, $2<\tau<2.5$, whatever the dimensionality
\citep{Kiang70a}. The exponents $\tau$ and $\sigma$, calculated from
the values of the thermodynamic exponents, are given in Table
\ref{tab1}.

\begin{table}[h!]
\begin{center}
\caption{\it Values of the thermodynamic critical exponents and of the
Fisher exponents $\sigma$ and $\tau$ at the critical point, in mean
field theory and in the universality classes of the Ising model in $2$
and $3$ dimensions (from \citealp{Gold92}).}
\label{tab1}
\begin{tabular}{lcccccc} \hline
             & $\alpha_T$ & $\beta_T$ & $\gamma_T$ & $\delta_T$ &
              $\sigma$ & $\tau$ \\ \hline
              
Mean field & 0 & $1/2$ & 1 & 3 & $2/3$ & $7/3$ \\ \hline

$2d-$Ising & 0 & $1/8$ & $7/4$ & 15 & $8/15 \simeq 0.53$ & $31/15 \simeq
2.07$ \\ \hline

$3d-$Ising & 0.12 & 0.325 & 1.24 & 4.8 & 0.64 & 2.21 \\ \hline

\end{tabular}
\end{center}
\end{table}

The link between thermodynamics and a geometric description of the
fluid stems from the perfect gas of clusters model. Indeed, Eqs.
(\ref{moka1})-(\ref{moka3}) of appendix \ref{sec:pgc} show that the
first moments of the cluster size distribution are equal to
thermodynamic quantities. For instance, moments of order $0$, $1$, and
$2$ behave respectively as the pressure, the density, and the
compressibility. Therefore, the second moment of the droplet size
distribution diverges like the compressibility at the critical point.

To test whether the droplet model is consistent with experimental
data, the exponent $\tau$ can be indirectly evaluated by means of the
compressibility factor $\beta P/\rho$, measured at the critical point
\citep{Kiang70a}. Indeed, from Eqs. (\ref{fpkt}) and (\ref{fro}), we
obtain for $x=y=1$
\begin{equation}
\label{comf}
\frac{P_{c}}{kT_{c}\rho_{c}}=\frac{\sum_{s=1}^{\infty}s^{-\tau}}{\sum_{s=1}^
{\infty}s^{1-\tau}}. 
\end{equation} 
For a large variety of real fluids, the value of $\tau$ inferred from
Eq. (\ref{comf}) is in the order of $2.2$, like the theoretical value
calculated from the thermodynamic exponents (see Table \ref{tab1}).
Moreover, a three-parameter fit of the compressibility factor along
the coexistence curve provides a $1\%$ agreement with experiment in
the gas phase for water and carbon dioxide \citep{Rath72}.

However, we have to point out that Fisher's model has some important
weaknesses:
\begin{itemize}
\item Fisher droplets are ``effective clusters'' which form a perfect
gas of clusters.\footnote{However, Stauffer and Kiang
\citeyearpar{Stau71a} extended the droplet model by taking into
account the excluded-volume and an additional attraction between the
clusters as a small perturbation. Critical exponents are not affected
by these corrections.} This basic assumption casts some doubt upon the
adequacy of the model at high density. Besides, as far as we know, no
attempt as been made to {\it observe} the Fisher droplets in a real
system.
\item The equation of state produces some singularities in the
supercritical phase, which is forbidden by the theorems of Yang and
Lee \citeyearpar{Yan-Lee52} \citep{Kiang70b}.\footnote{It is possible
to cancel out these singularities by adding some corrective terms,
without physical meaning, to the free energy of the clusters
\citep{Rea70}.}
\item The model does not describe the liquid phase beyond the
coexistence curve for $\rho>\rho_{cond}$ \citep{Kiang70b,Stau71b}.
\end{itemize}

Finally, despite these inadequacies, Fisher's phenomenological model
provides a first morphological interpretation of the critical point in
terms of clusters, as well as a thermodynamic study of the
condensation curve based on geometric arguments. During the past three
decades, Fisher droplets have become a guide for defining physical
clusters. Indeed, we have to {\it identify} clusters among the
particles of the fluid, in order to compare theoretical results with
experiments and numerical simulations. In brief, we need a microscopic
definition of clusters that behave like Fisher droplets at the
critical point.

\section{Microscopic definitions of clusters in the lattice-gas model}
\label{sec:lattice}

The power law behaviour of the Fisher droplet size distribution and
the divergence of the mean droplet size at the critical point remind
those of a percolation phase transition.\footnote{Stillinger Jr
\citeyearpar{Stil63} put forward the idea for understanding the
condensation as a percolation phase transition. In contrast, Fisher
does not mention percolation in his original article \citep{Fish67a}.}
Reader not familiar with percolation theory is referred to the book by
Stauffer and Aharony \citeyearpar{Stau94}. Hence, the works that
followed those of Fisher have searched for a microscopic definition of
clusters such that the thermodynamic critical point is a percolation
threshold characterized by the Ising critical exponents. In other
words, the microscopic clusters must verify the four following
criteria at the critical point \citep{Bind76a,Ker89,Stau90,Stau94}:
\begin{enumerate}
\item The mean cluster size distribution behaves as a power law:
$n_{s} \sim s^{-\tau}$ where $\tau$ is the Fisher exponent (see Table
\ref{tab1}).

\item A percolating cluster appears and its size $S_{max}$ varies like
the order parameter of the thermal phase transition, i.e the density
difference between liquid and gas (or the spontaneous magnetization in
the Ising model), with the exponent $\beta_{T}$: $S_{max} \sim
(T_c-T)^{\beta_{T}}$ for $T \le T_c $.

\item The mean cluster size, i.e the second moment of the cluster size
distribution\footnote{In percolation theory, the mean cluster size is
usually defined as $m_2/m_1$ where $m_k = \sum_{s} s^k n^{'}_s $ is
the moment of order $k$ of the cluster size distribution $n^{'}_s$
without the contribution of the largest cluster \citep{Stau94}.}
$S_{mean} \sim \sum_s s^2 n^{'}_s$, diverges as the compressibility (or
susceptibility) with the same exponent $\gamma_{T}$: $S_{mean} \sim
|T-T_c|^{-\gamma_{T}}$.

\item The connectedness length diverges as the thermal correlation
length with the same exponent $\nu_{T}$.
\end{enumerate}

We recall that in random percolation theory, the size of the largest
cluster, the mean cluster size, the connectedness length, and the
cluster size distribution are respectively described at the
percolation threshold by the following critical exponents $\beta_p$,
$\gamma_p$, $\nu_p$, $\sigma_p$ and $\tau_p$. These exponents are
universal for a given dimensionality and do not depend on the
particular process of percolation (site/bond percolation,
lattice/continuum space) \citep{Stau94}. For the sake of comparison,
the thermodynamic and random percolation exponents are respectively
given in Tables \ref{tab1} and \ref{tab2}.

\begin{table}[h!]
\begin{center}
\caption{\it Values of the critical exponents in mean field theory and
in the universality classes of random percolation theory in $2$ and
$3$ dimensions (from \citealp{Stau94}).}
\label{tab2}
\begin{tabular}{lcccc} \hline
             & $\beta_p$ & $\gamma_p$ & $\sigma_p$ &
             $\tau_p$ \\ \hline

Mean field  & 1 & 1 & 1/2 & 5/2 \\ \hline

$2d-$Percolation & 0.139 & 2.39 & 0.39 & 2.05 \\ \hline
 
$3d-$Percolation & 0.41 & 1.80 & 0.45 & 2.18 \\ \hline

\end{tabular}
\end{center}
\end{table}

\begin{figure}[h!]
\begin{center}
\includegraphics[angle=0,scale=.5]{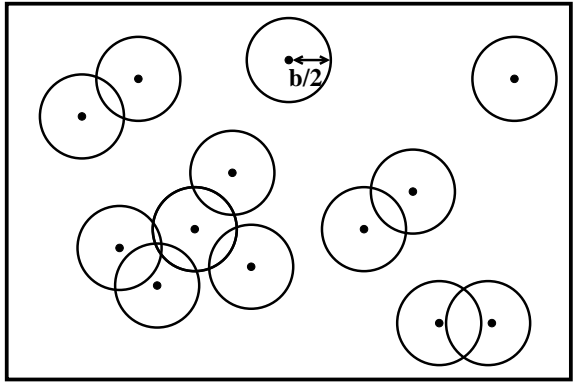}
\end{center}
\caption{\it Particles in a box, in two dimensions. The radius of the
discs centered at the positions of the particles is equal to $b/2$,
where $b$ is the cutoff distance. Configurational clusters are defined
by the overlap of these discs. Hence, in this figure, one can see two
monomers, three dimers and a cluster of size five.}
\label{fig2}
\end{figure}

The simplest and most natural way for defining and identifying
clusters among a group of particles is to base the criterion on
proximity in configurational space \citep{Band39a,Fish71}: two
particles are linked if the distance between them is shorter than a
given cutoff distance $b$. A cluster is then a group of particles
linked pair by pair\footnote{In the same spirit, a cluster can be
defined either as particles lying inside a spherical shell centered on
the center of mass of the cluster \citep{Reiss68,Lee73,Abra74,Abra75},
or as a density fluctuation exceeding a given threshold
\citep{Reiss90}.}  (see Fig. \ref{fig2}). In this article, such
clusters will be denoted ``configurational clusters''. We now have to
choose the value of the cutoff distance. At very low density, the
identification of clusters depends little on it. In contrast, this
definition becomes very sensitive to the choice of $b$ at higher
densities \citep{Stil63,Lee73,Abra75,Bind75}. As an illustration, if
we were to double the cutoff distance in Fig. \ref{fig2}, all the
particles would belong to the same cluster.

In order to apply the perfect gas of clusters model to a simple fluid,
we could choose $b \simeq constant \times r_{min}$, where $r_{min}$ is
the distance corresponding to the minimum of the potential
\citep{Bind75}. Nevertheless, when a realistic potential is used, the
attractive interaction between particles is never equal to zero, and
there is no natural cutoff distance to defining the clusters: the
choice of $b$ is arbitrary.

Thanks to its simplicity, the lattice-gas (Ising) model is a suitable
framework to study configurational clusters and to test the four
criteria of the Fisher droplets at the critical point.  As we shall
see in the next section, by choosing an attraction only between
particles occupying nearest neighbour sites of a lattice, it becomes
natural to choose a cutoff distance equal to the finite range of the
interaction \citep{Stoll72,Bind75,Mul76}.

\subsection{Ising clusters}

First, let us recall the definition of the lattice-gas model. The $M$
sites of a lattice are either empty or occupied by one of the $N$
particles. Particles occupying nearest neighbour sites interact with
an attractive energy $-\epsilon<0$. The Hamiltonian is then given by
$$
\mathcal{H}=\sum_{i=1}^M n_i\frac{p^2}{2m}-\epsilon \sum_{<i,j>} n_in_j
$$
where $n_i=0$ or 1, is the occupation number of the $i^{th}$
site. Mass conservation implies
$$
N=\sum_{i=1}^M n_{i}.
$$
This Hamiltonian can be considered as a simplification of the simple
fluid Hamiltonian given by Eq. (\ref{ea}). The equivalence of the
lattice-gas model and the Ising model allows us to use the framework
of the latter \citep{Lee-Yan52}. An up spin is assigned to an occupied
site, a down spin to an empty one, and a magnetic field $H$ allows for
the conservation (on average) of the number of particles. The
interaction energy between two nearest neighbour up spins is $J
=\epsilon/4$. The critical point is located at $T =T_c$, $\rho_c =0.5$
in the lattice-gas model,\footnote{By spin up/spin down or
particle/hole symmetry, the critical density of the lattice-gas model
equals to $0.5$ regardless of, either the structure of the lattice, or
the dimension of the configurational space.} and $H=0$ in the Ising
model. The line defined by $H=0$ and $T<T_c$ in the phase diagram
corresponds to the coexistence curve.

\begin{figure}[h!]
\begin{center}
\includegraphics[angle=0,scale=.5]{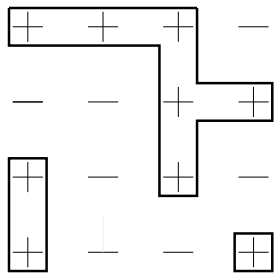}
\end{center}
\caption{\it Ising clusters on a square lattice. There is no
(diagonal) bond between the next-nearest neighbour spins up.}
\label{fig3}
\end{figure}

In the Ising model, configurational clusters are sets of nearest
neighbour sites occupied by an up spin. For a given configuration of
spins, the currently called ``Ising clusters'' are determined
uniquely, as is shown in Fig. \ref{fig3}. By definition, Ising
clusters are sets of interacting particles and there is no attraction
between them. However, it must be emphasized that these clusters have
an excluded volume which cannot be disregarded at high
density. Consequently, unlike the Fisher droplets, Ising clusters do
not make up a perfect gas of clusters.

The dimension $d$ of configurational space has a crucial effect on
thermal and geometric phase transitions. Therefore, the verification
of the four Fisher criteria has to be considered according to the
dimensionality.

In two dimensions: Coniglio and his collaborators
\citeyearpar{Coni77c} demonstrate by topological arguments that an
infinite cluster appears exactly at the critical point. Numerical
simulations on a square lattice exhibit a power law cluster size
distribution characterized by an exponent $\tau= 2.1 \pm 0.1$ in
agreement with Fisher's model \citep{Stoll72}. On the other hand, a
renormalization group calculation of exponent $\nu$ shows that the
average radius of the Ising clusters diverges like the correlation
length, that is $\nu=\nu_T$ \citep{Klein78}. The first and fourth
Fisher criteria are fulfilled by the Ising clusters.

In contrast, the exponents $\gamma$ and $\beta$, which give
respectively the critical behaviour of the mean cluster size and the
mean size of the largest cluster, do not have the expected values of
the thermodynamic exponents. Indeed, Sykes and Gaunt
\citeyearpar{Sykes76} show that $\gamma= 1.91 \pm 0.01$ (while
$\gamma_T = 1.75$) by means of series expansion calculations at low
temperature on a triangular lattice. Numerical simulations confirm
this disagreement: $\gamma = 1.901$ \citep{Fortu02} and $\beta = 0.052
\pm 0.030$ (while $\beta_T = 0.125$) \citep{Jan82}. In other words,
the second and third Fisher criteria are not verified. In two
dimensions Ising clusters do not behave like Fisher droplets.

In three dimensions, none of the four Fisher criteria are met. Along
the coexistence curve (for $H=0$), a percolating cluster appears at a
temperature lower than the critical one (at a density $\rho<\rho_c
=0.5$) \citep{Coni75}: at $T\simeq 0.94T_c$ and $\rho \simeq 0.25$ in
the cubic lattice from numerical simulations \citep{Mul74a,Mul74b},
and at $\rho \simeq 0.1$ in the face-centered cubic lattice from an
analytical calculation \citep{Sykes76}.

\begin{figure}[h!]
\begin{center}
\includegraphics[angle=0,scale=.35]{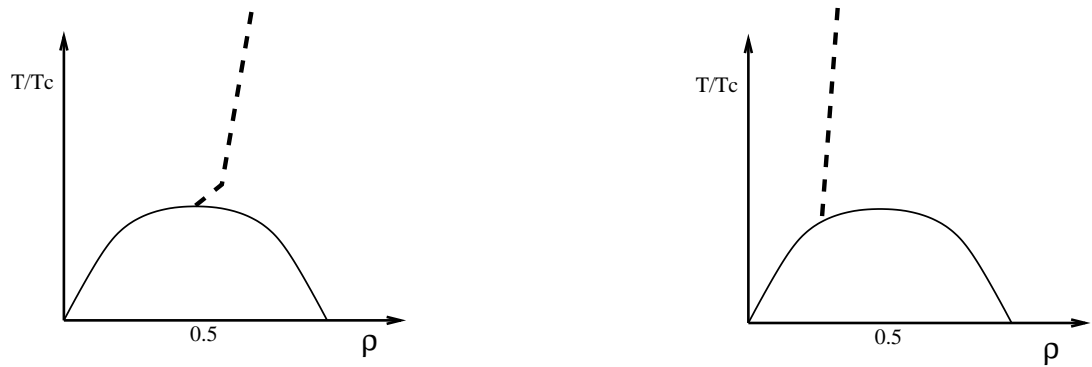}
\end{center}
\caption{\it Schematic phase diagram of the lattice-gas model. The
coexistence curve (solid line) and the Ising clusters percolation line
(dashed line) are represented for a square lattice ($q_c\simeq 0.59$)
on the left, and for a cubic lattice ($q_c\simeq 0.31$), on the
right. For a triangular lattice, $q_c=0.5$: the percolation line is
along the isochore $\rho=0.5$ (see text for notation).}
\label{fig4}
\end{figure}

Let us now consider the behaviour of the Ising clusters in the
supercritical phase. By using numerical simulations on a square
lattice, Odakaki and his coworkers \citeyearpar{Oda75} show that a
percolation line\footnote{In a previous work, Stillinger Jr
\citeyearpar{Stil63} had suggested the existence of a percolation line
in the supercritical phase.}  divides the supercritical region of the
$T-\rho$ phase diagram into two areas: at low density there are only
small size clusters, compared to the size of the system, and at high
density a macroscopic percolating cluster connects two opposite sides
of the lattice. This percolation line is plotted in Fig. \ref{fig4}
for two lattices. In two dimensions, and only in two dimensions, the
percolation line starts at the critical point. In the limit $T
\rightarrow \infty$, this line tends toward a density equal to the
random site percolation threshold $q_c$. Indeed, at high temperature,
interaction between spins becomes negligible compared to thermal
agitation. As a result, up spins (particles) are randomly distributed
with a probability equal to the density and we find the particular
case of the random site percolation \citep{Coni77c}. As we can see in
Fig. \ref{fig4}, the percolation behaviour is almost independent of
temperature. However, the positive slope shows that a short range
attraction favors the formation of clusters at low temperature,
compared to the random site percolation at infinite temperature.

Critical exponents that characterize the moments of the cluster size
distribution in the vicinity of the percolation line are equal to the
random percolation exponents. A calculation using the renormalization
group on a triangular lattice shows that the cluster radius diverges
with the exponent $\nu_p$ along the percolation line
\citep{Klein78}. On the other hand, numerical simulations provide
$\beta = 0.145 \pm 0.027$ (whereas $\beta_p=0.139$) \citep{Oda75}.

In an infinite Bethe lattice (mean field theory), Coniglio
\citeyearpar{Coni75,Coni76} demonstrates that the percolation line
does not start at the critical point, but at a lower density than the
critical one, like in three-dimension lattices.

In conclusion, Ising clusters do not behave like Fisher droplets at
the thermodynamic critical point, regardless of the
dimensionality. Moreover, Ising clusters give rise to a percolation
line in the supercritical phase. It is important to remark that the
presence of a percolating Ising cluster in the supercritical phase was
considered as non-physical, since on the basis of Fisher's model, it
would imply thermodynamic singularities \citep{Rea70,Bind75,Mul76}. We
shall return to this point later, when we discuss Swendsen-Wang
clusters (see section \ref{sec:sw}). Since Ising clusters are too
large to behave like Fisher droplets, smaller clusters have to be
defined \citep{Bind76a}.

\subsection{Coniglio-Klein clusters}
\label{sec:ck}

Fortuin and Kasteleyn \citep{Kast69,Fort72} propose a correspondence
between thermodynamic and geometric quantities through a formalism
based on graph theory within the framework of the Ising model. When
the probability that two up spins are linked is properly chosen, the
thermodynamic critical point is exactly a geometric critical point
characterized by the exponents of the universality class of the Ising
model.

\begin{figure}[h!]
\begin{center}
\includegraphics[angle=0,scale=.5]{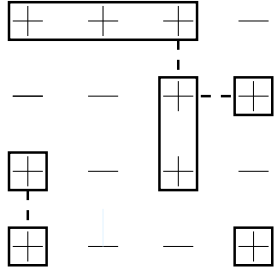}
\end{center}
\caption{\it A configuration of Coniglio-Klein clusters on a square
lattice obtained from the Ising clusters of Fig. \ref{fig3}. The
broken bonds between nearest neighbour up spins are represented by
dashed lines.}
\label{fig5}
\end{figure}

Using an Hamiltonian formulation of random percolation based on an
analytical extension of the Potts model \citep{Wu82} with $x$ colors
when $x \rightarrow 1$, Coniglio and Klein \citeyearpar{Coni80} obtain
the same results as Fortuin and Kasteleyn, and define the so-called
``Coniglio-Klein clusters''. For a discussion about the relation
between the approaches proposed by Fortuin and Kasteleyn and by
Coniglio and Klein, we refer the reader to \citep{Hu84} and to the
Appendix of \citep{Coni01}. Two nearest neighbour sites occupied by an
up spin are connected by a bond with the probability\footnote{ We
choose to express $p_{ck}$ in the lattice-gas framework. Because
$J=\epsilon/4$, we have in the Ising model: $p_{ck}=1-e^{-2\beta J}$.}
\begin{equation}
\label{pck} 
p_{ck}(\beta \epsilon)=1-e^{-\frac{\beta \epsilon}{2}}. 
\end{equation}
It is sometimes helpful to think of $1-p_{ck}$ as the probability of
breaking a bond between two nearest neighbouring up spins in an Ising
cluster. Coniglio-Klein clusters are defined as sets of up spins
linked by bonds.\footnote{Following an idea of Binder
\citeyearpar{Bind76a}, Alexandrowicz \citeyearpar{Alex88,Alex89}
proposes to define clusters as sets of up (or down) spins surrounded
by a perimeter of up and down spins at their average proportion. This
leads to the same fractal dimension as the Coniglio-Klein's definition
at the critical point.}  We remark that the probability $p_{ck}$ is
independent of the dimensionality. Figure \ref{fig5} shows a
configuration of Coniglio-Klein clusters obtained from the Ising
clusters represented in Fig. \ref{fig3}. It must be emphasized that
unlike Fisher droplets, Coniglio-Klein clusters do not form a perfect
gas: besides having an hard-core repulsion like Ising clusters, they
interact through nearest neighbouring up spins that are not connected
by a bond, as illustrated in Fig. \ref{fig5}.

In spite of this crucial difference, Coniglio-Klein clusters satisfy
the four Fisher criteria at the critical point
\citep{Coni80}. Numerical simulations confirm this result and give for
$H=0$ and $T=T_{c}$, values of the geometric exponents equal to the
thermodynamic ones: in two dimensions, an infinite cluster appears at
the critical point \citep{Otta81}, $\gamma=1.77 \pm 0.03$ (while
$\gamma_{T}=1.75$), $\beta \simeq 0.125= \beta_{T}$ \citep{Jan82}, and
$\tau=2.04$ (while the Fisher exponent is $\tau=2.07$)
\citep{Liverpool96}. In three dimensions, Coniglio-Klein clusters
diverge at the critical point \citep{Stau81} with $\gamma \simeq 1.2$
(while $\gamma_{T}=1.24$), and $\tau \simeq 2.25$ (while the Fisher
exponent is $\tau=2.21$) \citep{Rouss82}.

\begin{figure}[h!]
\begin{center}
\includegraphics[angle=0,scale=.5]{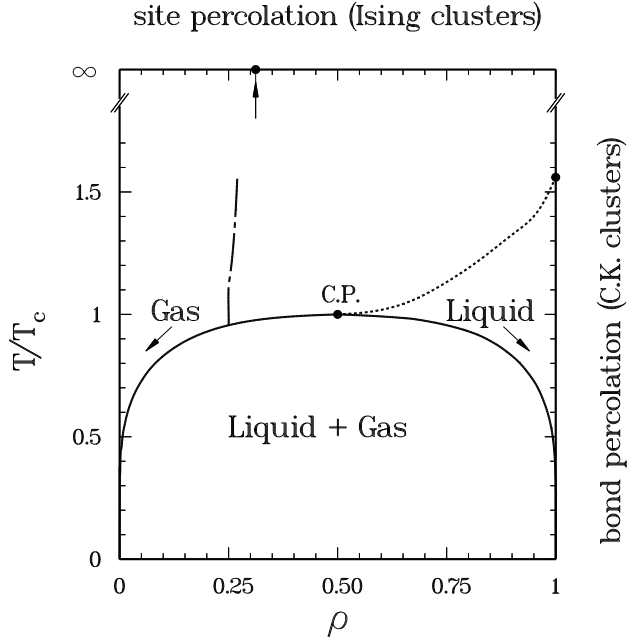}
\end{center}
\caption{\it Phase diagram $T-\rho$ of the lattice-gas model
determined numerically in a simple cubic lattice. Coexistence curve
(solid line). Critical point (C.P.). The Ising clusters percolation
line (dashed-dotted line) ends at the site percolation threshold
($q_c\simeq 0.31$) when $T \rightarrow \infty$. The Coniglio-Klein
percolation line, or Kert\'esz line (dotted line), ends at the bond
percolation threshold when $\rho=1$ (from \citealp{Campi97}).}
\label{fig6}
\end{figure}

Coniglio-Klein clusters give rise to a new percolation line, known as
the ``Kert\'esz line'' or ``Coniglio-Klein line'', which is shown in
Fig. \ref{fig6}. As it should be, this line starts at the critical
point ($\rho_c=0.5$, $H=0$) whatever the dimensionality, and reaches
the point ($\rho=1$, $H=\infty$) at temperature $T_{b_c}$ given by
$1-e^{-\frac{\epsilon}{2kT_{b_c}}}=p_{b_c}$, where $p_{b_c}$ is the
random bond percolation threshold. Indeed, for $\rho=1$, all the sites
are occupied by an up spin (particle) and we find the particular case
of random bond percolation \citep{Heer81}.

As we can see in Fig. \ref{fig6}, the supercritical area of the phase
diagram is divided into a non-percolating region at high temperature
and low density, and a percolating region at low temperature and high
density. Along the Kert\'esz line, geometric exponents are equal to
those of random percolation \citep{Coni80,Stau81}. Therefore, values
of geometric exponents pass discontinuously from the thermodynamic
ones at the critical point, to the random percolation values along the
line, for $T>T_{c}$ and $\rho> \rho_c$ ($H>0$).

It is interesting to remark that when interactions with next or second
neighbours are taken into account, the critical point remains a
geometric critical point for Coniglio-Klein clusters\footnote{For
Ising clusters, the density of the geometric critical point along the
coexistence curve obviously decreases as the range of attraction
increases \citep{Bug85}.} \citep{Jan82}. This result suggests that
moving on to the continuum limit and considering a more realistic
potential with a larger range of attraction, the critical point is a
percolation threshold.\footnote{As a matter of fact, Fortuin and
Kasteleyn's formalism has been generalized to the case of a short
range potential in a continuous space \citep{Dro96}.} This
possibility, as well as the existence of a percolation line in a real
fluid, will be discuss later on in section \ref{sec:self}.

To close this subsection, we briefly discuss the behaviour of
Coniglio-Klein and Ising clusters below the coexistence curve. At {\it
thermal equilibrium}, Monte-Carlo simulations in three dimensions show
that no critical percolation behaviour exists when phase separation
occurs: critical percolation lines do not get into the two-phase
region \citep{Campi-Kri02}. Nevertheless, below the coexistence curve,
there is always a macroscopic cluster (either a Coniglio-Klein or an
Ising cluster), in the sense that a finite fraction of the particles
belongs to the largest cluster.\footnote{To put it more precisely, the
macroscopic cluster can be either too compact to span the system or
enough spread out to percolate, according to temperature and density
values. In the last case, although the largest cluster connects two
opposite sizes of the lattice, we recall that no critical percolation
transition is observed, the distribution of finite size clusters
remains exponentially decreasing and does not display any power law
behaviour \citep{Campi-Kri02}.} This is not the purpose of this review
article to deal with the out of equilibrium and nucleation properties
of simple fluids, however, we mention that Coniglio-Klein clusters, as
well as Ising clusters, fulfill the phenomenological laws of Classical
Nucleation Theory below the coexistence curve \citep{Heer84,Ray90}. On
the other hand, in mean field theories, the compressibility diverges
not only at the critical point but also along the well-defined
spinodal line which separates the metastable and unstable regions. The
mapping between thermodynamics and geometry in terms of Coniglio-Klein
clusters cannot be extended into the metastable region
\citep{Schio98}. However, by introducing the bond probability $P_B =1
-e^{-\beta \epsilon(1-\rho)}$ between particles, the mean size of
these new clusters diverges along the mean field spinodal line
\citep{Heer84,Klein90}.

\subsection{Swendsen-Wang clusters}
\label{sec:sw}

As we have seen in the last section, Coniglio-Klein clusters behave
like Fisher droplets at the critical point ($T=T_c$, $H=0$). However,
the percolation line in the supercritical phase ($T>T_c$, $H>0$) was
interpreted as a non-physical manifestation of these clusters. Within
the context of the Fisher droplet model (or more generally in the
perfect gas of clusters model), a percolation behaviour should imply
singularities of thermodynamic quantities, which is forbidden in the
supercritical phase by the theorems of Yang and Lee
\citeyearpar{Yan-Lee52}. The aim was then to define clusters that have
a geometric critical behaviour (percolation threshold) at, {\it and
only at}, the thermodynamic critical point. In this spirit, the
Kert\'esz line was rather cumbersome \citep{Coni89,Ker89,Stau90}.

\begin{figure}[h!]
\begin{center}
\includegraphics[angle=0,scale=.4]{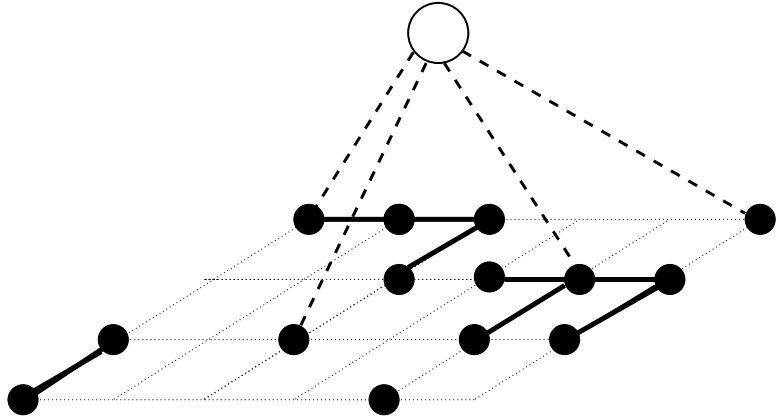}
\end{center}
\caption{\it Up Spins (particles) are represented in the lattice-gas
formulation by a black dot. The white dot is the ``ghost spin''.
Bonds between up spins are active with probability
$p_{ck}=1-e^{-2\beta J}$ (heavy line) and bonds between up spins and
the ``ghost spin'' are active with probability $p_{h}=1-e^{-2\beta H}$
(dashed line). Some up spins are connected only through the ``ghost
spin''.}
\label{fig7}
\end{figure}

In order to get rid of the percolation line, Coniglio and his
collaborators \citeyearpar{Coni89}, as well as Wang
\citeyearpar{Wang89}, introduce a definition of clusters based on the
work of Fortuin and Kasteleyn \citep{Kast69,Fort72}, and Swendsen and
Wang\footnote{Swendsen-Wang algorithm, which allows one to reduce the
critical slowing down at the critical point, is essentially based on
the Fortuin and Kasteleyn's formalism
\citep{Swen87,Wang90,Liverpool96}.}  \citeyearpar{Swen87}. These
so-called ``Swendsen-Wang clusters'' are defined like the
Coniglio-Klein clusters, but in the presence of a positive (negative)
magnetic field, $H$, up (down) spins are in addition connected to a
``ghost spin'' with the probability $p_h=1-e^{-\mid h \mid}$, where
$h=2\beta H$. This model is best explained by referring to
Fig. \ref{fig7}.  At each point of the phase diagram, the cluster {\it
finite} size distributions of Swendsen-Wang clusters $n_{s}^{sw}$, and
Coniglio-Klein cluster $n_{s}^{ck}$, are related by \citep{Wang89}:
$$
n_{s}^{sw}=2n_{s}^{ck}e^{-\mid h \mid s} \qquad \textrm{where $s$ is
finite}.
$$
Hence, as soon as $H$ is different from zero, a percolating cluster,
which contains the ghost site, is present in the system whatever the
temperature. Indeed, for $H>0$, even if the magnetic field is weak, a
macroscopic fraction ($1-e^{-2\beta H}\simeq 2\beta H$) of up spins is
connected to the ``ghost spin''. As a result, there is a percolating
cluster at each point of the phase diagram located above the
coexistence curve \citep{Stau90,Adler91}. In the thermodynamic limit,
any cluster that contains the ``ghost spin'' is infinite, and vice
versa. The infinite cluster is unique. {\it Finite} Swendsen-Wang
clusters behave as we expected: they verify the four Fisher criteria
at the critical point, as the Coniglio-Klein clusters do. Besides,
their mean size does not diverge in the supercritical phase. The
percolation line is eliminated.

However, even if the finite Swendsen-Wang cluster size distribution
does not have a power law behaviour in the supercritical phase,
something happens along the Kert\'esz line. Wang \citeyearpar{Wang89}
calculated $n_{s}^{sw}$, for large but finite clusters, by means of
numerical simulations. According to the region of the phase diagram
(see Fig. \ref{fig6}), he found:
\begin{itemize}
\item Above the Kert\'esz line, $n_{s}^{sw} \sim e^{-c_1s -\mid h \mid
s }$
\item Along the Kert\'esz line, $n_{s}^{sw} \sim s^{-\tau}e^{-\mid h \mid s}$
\item Below the Kert\'esz line, $n_{s}^{sw} \sim
e^{-c_2s^{\frac{2}{3}}-\mid h\mid s}$
\end{itemize}
where the coefficients $c_1$ and $c_2$ depend on temperature and
magnetic field. The Swendsen-Wang cluster size distribution decreases
exponentially everywhere in the supercritical phase. However, the
``surface term'' of $n_{s}^{sw}$, varying like $s^{\frac{2}{3}}$ below
the line, goes to zero along and above the line. Kert\'esz
\citeyearpar{Ker89} and Stauffer \citeyearpar{Stau90} suggested that
the cancellation of the surface term in the expression for the cluster
size distribution should imply a weak singularity of the cluster free
energy.\footnote{According to Kert\'esz \citeyearpar{Ker89} the
contribution of the percolating cluster to the total free energy
should cancel out this weak singularity and the partition function
would be of course analytical.}  Discussing this result, Stauffer and
Aharony \citeyearpar{Stau94} wrote: ``If correct, it would mean that
we have to make more precise the century-old wisdom that nothing
happens if we move continuously from the vapor to the liquid phase of
a fluid by heating it above the critical temperature.''

At the beginning of the nineties, this claim sounded
conclusive. Nevertheless, we believe this interpretation of the
Kert\'esz line to be incorrect and propose a new point of
view. Indeed, it must be emphasized that this argument is based on the
hypotheses of the perfect gas of clusters in which (see
Eq. (\ref{nsf}))
\begin{equation}
\label{nspg}
n_s=q_s z^{s}=e^{-\beta F_s+s\ln z}
\end{equation}
Remember that Coniglio-Klein clusters, as well as Swendsen-Wang
clusters, do not form a perfect gas of clusters. We can see therefore
no valid reason for writing their cluster size distributions as
Eq. (\ref{nspg}). In other words, the particular behaviour of the
Coniglio-Klein and Swendsen-Wang cluster size distributions along the
Kert\'esz line does not imply even weak thermodynamic
singularities. Furthermore, as far as we know there is no physical
interpretation of the ghost spin and Swendsen-Wang clusters. We shall
see that this is not the case for Coniglio-Klein clusters.

Before we move on to the next section, we point out a general fact
about clustering in simple fluids: the definition of clusters does
depend on the physical problem we want to study. For example, Ising
clusters do not provide a geometric interpretation of the liquid-gas
transition. However, properties of electrical conductivity can be
understood in terms of these clusters of interacting particles
\citep{Oda75,Yang98}. Moreover, in the continuum case, configurational
clusters have been used to describe the dynamics of hydrogen bonds in
liquid water \citep{Luzar96,Starr99}.

\section{Self-bound clusters}
\label{sec:self}

From now on, our purpose is to define clusters as self-bound sets of
particles. A typical physical situation in which it is relevant to
introduce ``self-bound clusters'' is the fragmentation of a piece of
matter (such as an atomic nucleus, an atomic aggregate, or a liquid
droplet) \citep{Campi01a}. In particular, a question of interest is
the existence of {\it physical} clusters in a dense medium, before
fragmentation occurs. This topic seems to contrast strongly with the
discussion of the last sections. Yet, we shall see how such clusters
can shed some light on the geometric interpretation of the liquid-gas
phase transition.

Self-bound clusters would be stable if they were isolated from each
other. In a dense medium, their definition cannot rest only on a
criterion of proximity in configurational space. Such a definition
would make no allowance for the relative motion of the particles. At
high temperature, two particles that strongly interact are close to
each other at a given time. Because of their high velocities, they may
quickly move away from each other. A definition of clusters that takes
into account the relative velocities of the particles corrects this
inadequacy.\footnote{An alternative approach is to introduce an
arbitrary bond life time, in addition to the cutoff distance: two
particles are linked if they remain within a certain distance during a
given time \citep{Pugna00}. However, not one but two adjustable
parameters have to be chosen.}

In section \ref{defself}, we present two definitions based on
energetic criteria which allow us to identify self-bound clusters in a
fluid knowing the positions and velocities of the particles. The first
definition is global and based on minimizing the interaction energy
between clusters. The second one is a local energetic criterion of
bond activation. Next, we show that Coniglio-Klein clusters are
self-bound on average. This crucial link between thermodynamics and
the morphology of a fluid in the lattice-gas model will provide a
geometric interpretation of the liquid-gas phase transition in terms
of {\it physical} clusters. This point will be developed in section
\ref{these} for a Lennard-Jones fluid.

\subsection{Definitions of self-bound clusters}
\label{defself}

\subsubsection{A global criterion}

At very low density, clusters are naturally defined as sets of
particles that strongly interact with each other and do not interact
(or only very loosely) with particles belonging to other 
clusters. This is the domain of validity of the perfect gas of
clusters model.

In a dense fluid, by an argument of continuity, self-bound clusters
can be defined by the partition of particles that minimizes the
interaction energy (in absolute value) between clusters, or
equivalently that minimizes the sum of the clusters internal
energies. Indeed, the total energy $E$ of the system divided in an
arbitrary partition of clusters can be written
\begin{equation} 
\label{parti}
E= \sum_{C} \frac{1}{2}mN_{C} V_{C}^2 + \sum_{C} U_{C} +\sum_{C,C'}
V_{int}(C,C')
\end{equation} 
where $\sum_{C}$ is a sum over all the clusters, and $N_C$, $U_C$, and
$V_C$ are respectively the number of particles, the internal energy,
and the velocity of the center of mass of cluster $C$. The term
$V_{int}(C,C')$ is the interaction energy between clusters $C$ and
$C^{'}$. We then have to find the partition of clusters that minimizes
\begin{equation} 
\label{ranec} 
U(P)= \sum_{C} U_{C}
\end{equation}
This minimization problem has an exact, but not necessarily unique,
solution. When the number of particles is small enough to enumerate
all the possible partitions, a solution can be found
easily. Otherwise, we can use two different methods for finding an
optimal solution.

The first one was proposed by Dorso and Randrup
\citeyearpar{Dorso93}. It is based on an optimization procedure called
``Simulated Annealing'' \citep{Kirk84}: consider an arbitrary
partition of clusters $P$, a particle is randomly chosen and
transferred from its original cluster $C$ to another cluster $C'$. Of
course, the particle remains at the same position in the fluid but we
decide that it now belongs to another cluster.  We obtain a new
partition $P^{'}$. The variation of the total internal energy between
the two partitions is
$$
\Delta U= U(P^{'})-U(P)=U'_{C}+U'_{C'}-U_{C}-U_{C'}
$$
where $U'_C$ and $U'_{C'}$ are respectively the internal energies of
clusters $C$ and $C'$ after the particle has been transferred. If
$\Delta U<0$, the partition $P^{'}$ is accepted. If $\Delta U>0$, the
partition $P^{'}$ is accepted with the probability $e^{-\frac{\Delta
U}{T_{e}}}$, where $T_{e}$ is a control parameter. The value of $T_e$
is progressively decreased until there is no more modification of the
cluster partition on a given number of attempts. As usual with
Metropolis methods, the total internal energy of the final partition
may be only a local minimum.

A second method, denoted ``BFM`` (for Binary Fusion Method) has been
developed by Puente \citeyearpar{Puente99}: initially, we suppose that
there are only monomers in the fluid.  The multiplicity, which is the
number of clusters, is maximal, that is $m_{0}=N$. Then, at each
iteration we merge the two clusters that provide the lowest value of
the function $U(P)$ (see Eq. (\ref{ranec})). The multiplicity is then
reduced by one at each iteration. Therefore, at the first iteration a
dimer is built, at the second either another dimer or a trimer, and so
on. The process stops when the total internal energy cannot be reduced
any more. This deterministic method follows a certain trajectory in
the phase space of the partitions of clusters that does not
necessarily lead to the optimal partition.  On the other hand, the
Simulated Annealing method can probe the phase space in a
probabilistic way, which depends on arbitrary chosen parameters (like
the number of Metropolis trials or the way $T_e$ decreases). According
to their authors, these two procedures provide clusters that are
stable against particle emission. The definition of stability against
particle emission is presented at the beginning of appendix
\ref{stabhill}.

By means of numerical simulations of a Lennard-Jones fluid, Puente
\citeyearpar{Puente99} has shown that the Simulated Annealing and the
BFM methods produce very similar cluster size distributions, but the
last one is 10-20 times faster for the system considered. However, it
has to be noted that both of these methods are very time consuming and
cannot be used in practice for large systems made up of more than
(say) $N=1000$ particles.

\subsubsection{A local criterion: Hill clusters}
\label{hill}

The definition of clusters proposed by Hill \citeyearpar{Hill55} rests
on a bond activation between pairs of particles. At a given time, two
particles $i$ and $j$, of respective velocities $\vec v_{i}$ and $\vec
v_{j}$, are linked if their relative kinetic energy $K_{r}$ is lower
than the negative of their attractive interaction energy $u(r_{ij})$:
\begin{equation}
\qquad K_{r}=\frac{m}{4}(\vec v_{i}-\vec v_{j})^2 \le -u(r_{ij})
\Rightarrow \textrm{$i$ and $j$ are linked}.
\label{crihill}
\end{equation}
In other words, there is a bond between two particles if they are
close enough to each other in the phase space, and not only in the
configurational space. A Hill cluster is by definition a set of
particles connected by bonds two by two. When the positions and
velocities of the particles are known, for instance by means of
molecular dynamics simulations \citep{Allen87}, this local definition
is completely deterministic and does not depend on adjustable
parameters. It has to be noticed that the Hill's criterion can be used
as well with Monte-Carlo calculations, although the velocities of the
particles are not provided. Indeed, we checked by means of molecular
dynamics simulations that mean cluster size distributions are not
modified when we replace the velocity of each particle, provided by
molecular dynamics simulations, with velocity components taken at
random from a Gaussian distribution characterized by the temperature
of the system \citep{Sat00}. As far as mean cluster size distributions
are concerned, position-velocity correlations are irrelevant and
Monte-Carlo calculations can be used by allocating to each particle a
velocity from a Gaussian distribution. Of course, time dependent
properties and dynamics of the self-bound clusters are investigated
only by molecular dynamics simulations.

A question that naturally arises at this point is the comparison
between the global (Simulated Annealing and BFM) and the local (Hill)
criterion. Performing molecular dynamics calculations with a
Lennard-Jones fluid, Campi and his collaborators
\citep{Campi01a,Sat00} have found that the cluster size distributions
obtained with these three definitions are very similar. To quote Hill
\citeyearpar{Hill55}: ''A cluster is [\dots] a fairly definite
physical concept so that all physically reasonable definitions should
lead to rather similar predictions of the cluster size distribution.''
However, for a system of $N=1000$ particles, Hill's criterion is 300
times faster to implement and can be used in practice for much larger
systems, as we shall see in section \ref{these}.

Because it is arduous to carry out an analytical approach by using the
Hill's criterion, it may be relevant to deal with a probabilistic
formulation of the criterion (\ref{crihill}). That requires us to
calculate the probability that two particles are linked.

In a fluid at equilibrium, velocity components of the particles are
given by a Gaussian distribution characterized by a variance $1/(\beta
m)$, where $m$ is the mass of the particles and $\beta=1/kT$. The
criterion (\ref{crihill}) involves the relative kinetic energy between
particles $i$ and $j$ which can be seen as the kinetic energy of a
fictitious particle of reduced mass $\mu=m/2$ and relative velocity
$\vec v_r = \vec v_i -\vec v_j$. The velocity components of the
fictitious particle are then given by a Gaussian distribution with
variance $2/\beta \mu=1/ \beta m$.

The probability $p_{Hill}(r,\beta)$ that two particles separated by a
distance $r$ are linked is then the probability that
$K_r=\mu v_{r}^2 /2 \le -u(r)$, that is  
\begin{eqnarray*}  
p_{Hill}(r,\beta) &=&\frac{\int_{0}^{\sqrt{-2u(r)/\mu}}
v_r^2e^{-\frac{\beta}{2}\mu v_{r}^2}dv_r}{\int_{0}^{\infty}
v_r^2e^{-\frac{\beta}{2}\mu v_{r}^2}dv_r}\\ &=&
\frac{\int_{0}^{\sqrt{-\beta u(r)}} x^2e^{-x^2}dx}{\int_{0}^{\infty}
x^2e^{-x^2}dx} \\ &=& \frac{\Gamma (\frac{3}{2},-\beta
u(r))}{\Gamma(\frac{3}{2})}
\end{eqnarray*}
where $\Gamma(n,x)$ is the incomplete gamma function. In this
probabilistic framework, bonds are independent of each other, like in
a random percolation process. A a consequence, we lose the velocity
correlations between particles which are central in the deterministic
criterion. We shall see in section \ref{these} that clusters defined
with the deterministic definition are much larger than the
probabilistic clusters.

By using the probability $p_{Hill}(r,\beta)$, Hill
\citeyearpar{Hill55} introduced two effective potentials between bound
and unbound pairs of particles for expressing the mean number of
clusters as density and activity series expansions.\footnote{This
method, which takes into account the interactions between clusters,
can be formally applied to any microscopic definition of clusters.} A
very similar approach has been used recently to calculate the first
order corrections (namely by considering the interaction between a
monomer and a self-bound cluster) to the perfect gas of clusters at
low density \citep{Soto97}. Coniglio and his collaborators
\citeyearpar{Coni77a,Coni77b} extended the Hill's method and developed
an integral equation theory of the pair connectedness function, which
is proportional to the probability that two particles belong to the
same cluster.\footnote{For a comprehensive review about integral
equation theory, see \citep{Caccamo96}.}  With an
Ornstein-Zernike-type equation and a closure relation, like the
Percus-Yevick approximation, the mean cluster size can be obtained
from the integration of the pair connectedness function. Very
recently, a generalization to the case of the deterministic Hill's
criterion has been proposed and should provide interesting results
\citep{Pugna00,Pugna02}.

\subsubsection{Coniglio-Klein and self-bound clusters}
\label{hillself}

In the lattice-gas model, interaction is equal to $-\epsilon$ between
nearest neighbour particles, and $0$ otherwise. The probability that
two nearest neighbour particles are linked becomes \citep{Campi97}
\begin{eqnarray}
p_{Hill}(\beta\epsilon) &=& \frac{4}{\sqrt{\pi}}
\int_{0}^{\sqrt{\beta\epsilon}} e^{-x^2}x^2dx \nonumber \\ 
&=& 1-\frac{4}{\sqrt{\pi}}\int_{\sqrt{\beta\epsilon}} ^{\infty}
x^2e^{-x^2}dx. \label{hillprob}
\end{eqnarray}
We show in appendix \ref{stabhill} that probabilistic Hill clusters
are stable on average against particle emission. By performing a
Laguerre expansion of Eq. (\ref{hillprob}), Campi and Krivine
\citeyearpar{Campi97} have found
\begin{eqnarray}
p_{Hill}(\beta\epsilon) &=& 1-0.911e^{-\frac{\beta\epsilon}{2}}-0.177(1-\beta
\epsilon)e^{-\frac{\beta\epsilon}{2}} +\dots \nonumber \\
&=&p_{ck}(\beta\epsilon)-0.088(1-2.01\beta\epsilon)e^{-\frac{\beta\epsilon}{2}}
+\dots  \nonumber \\
&\simeq& p_{ck}(\beta\epsilon)  \label{hilck}
\end{eqnarray}
where $p_{ck}(\beta\epsilon)=1-e^{-\frac{\beta\epsilon}{2}}$ is the
probability introduced by Coniglio and Klein (see section
\ref{sec:ck}). Equation (\ref{hilck}) shows that probabilities
$p_{ck}$ and $p_{Hill}$ are very close to each other (for example at
the critical point of the cubic lattice, the relative error between
the two probabilities is less than $6\%$).\footnote{Pan and Das Gupta
\citeyearpar{Dasgupta95} had already noticed this similarity using
numerical simulations.} This is a crucial result: Coniglio-Klein
clusters, which have been introduced specifically to provide a
correspondence between geometric and thermodynamic critical behaviour,
do have a physical interpretation in terms of self-bound
clusters. Conversely, self-bound clusters give rise to a percolation
line which starts at the critical point in the framework of the
lattice-gas model.

\subsection{Percolation lines in simple fluids}
\label{these}

\begin{figure}[h!]
\begin{center}
\includegraphics[angle=-90,scale=.5]{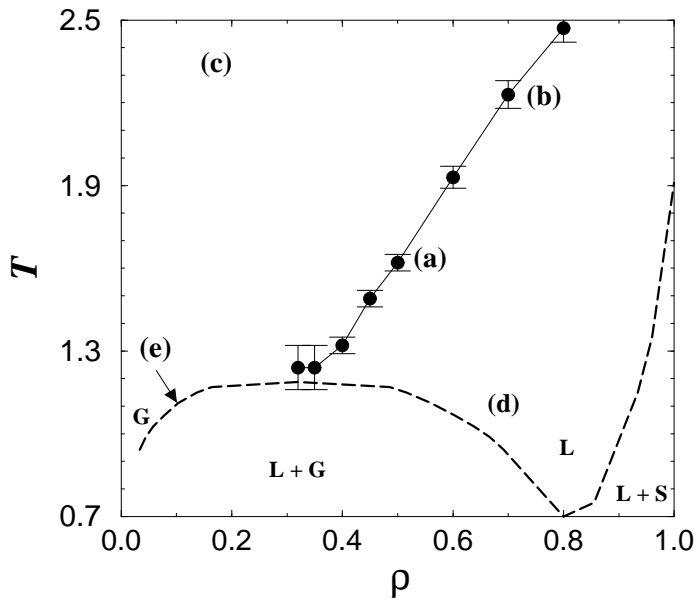}
\end{center}
\caption{\it Phase diagram of the Lennard-Jones fluid. The dashed line
indicates the coexistence curves between the gas (G), liquid (L) and
solid phases (S). The continuous line is the critical percolation line
of self-bound clusters. Temperature $T$ and density $\rho$ are in
units of $\epsilon$ and $\sigma^{-3}$ respectively (see
Eq. (\ref{lj})) (from \citealp{Campi01a}).}
\label{fig8}
\end{figure}

Numerical simulations of a Lennard-Jones fluid provide a more
realistic theoretical framework than the lattice-gas model. With the
help of canonical Monte Carlo and microcanonical molecular dynamics
calculations \citep{Allen87}, Campi and his collaborators
\citep{Sat00,Campi01a} have studied systems made up of a large number
of particles ($N = 11664$) interacting through a Lennard-Jones
potential given by Eq. (\ref{lj}). The phase diagram of this simple
fluid is presented in Fig. \ref{fig8}. It must be emphasized that the
phase diagram, as well as the results we present below, are
qualitatively insensitive to this particular choice of potential and
can be considered as generic for simple fluids. To reduce the computer
time calculation, self-bound clusters are recognized by using the
local deterministic criterion proposed by Hill (see section
\ref{hill}).

\begin{figure}[h!]
\begin{center}
\includegraphics[angle=0,scale=.7]{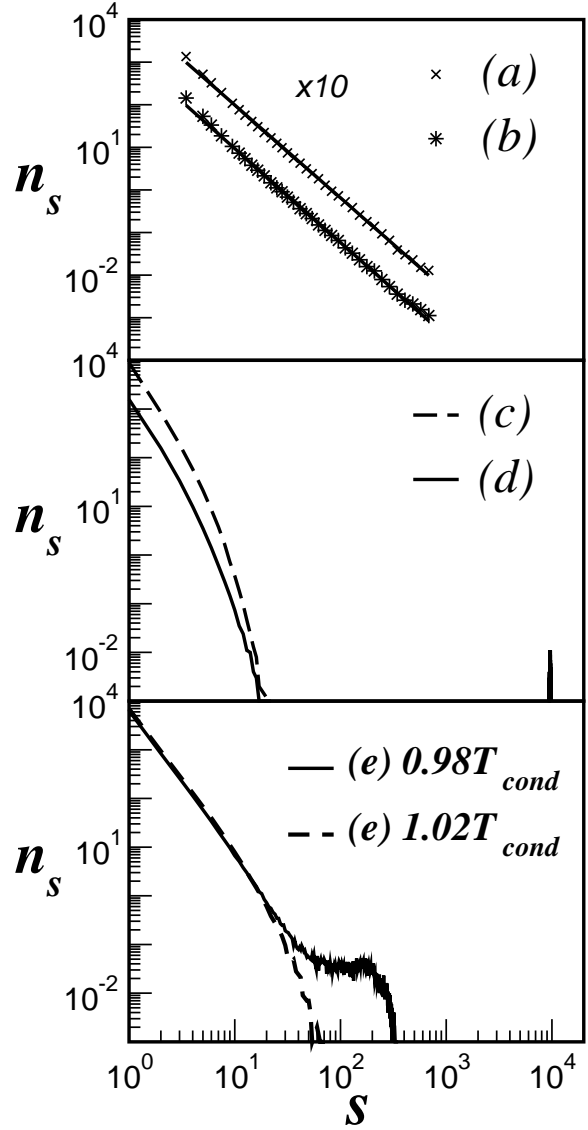}
\end{center}
\caption{\it Cluster size distributions $n_s$ at points (a)-(b),
(c)-(d) and (e) of the phase diagram (see Fig. \ref{fig8}). For curve
(d): the contribution of the percolating cluster is sharply peaked
around $s \simeq 9500$. For curves (e): $T_{cond}$ is the temperature
of condensation at $\rho=0.1$ (from \citealp{Campi01a}).}
\label{fig9}
\end{figure}

As a first result, a ``macroscopic'' cluster appears as soon as the
liquid-gas phase transition takes place. Figure \ref{fig9} represents
two cluster size distributions plotted around the point $(e)$ of the
phase diagram (see Fig. \ref{fig8}), just above and just below the
coexistence curve. We can see a drastic change in the mean number of
large clusters which corresponds very well to the crossing of the
coexistence curve. Obtained without any free parameter, this clean
correspondence between thermodynamic and morphological changes cannot
but support the physical nature of these clusters. It is important to
note that along the coexistence curve, the cluster size distribution
is not a power law and therefore, does not present a geometric
(percolation) critical behaviour.

What is more, self-bound clusters give birth to a critical percolation
line\footnote{This line should not be confused with the Fisher-Widom
line \citep{Fish69}, which divides the supercritical region into a
gas-like and a liquid-like domain (with probably no relevance to
clustering) according to the behaviour of the radial distribution
function, nor with the extrapolation of the rectilinear diameter line
\citep{Nishi00}, which is the locus of points of maximum
compressibility.} represented in Fig. \ref{fig8}. To illustrate this
percolation behaviour, the cluster size distribution is plotted in
Fig. \ref{fig9} for several points of the phase diagram.  Along the
line the cluster size distribution behaves as a power law
characterized by an exponent $\tau=2.20\pm0.05$ (while in random
percolation theory $\tau_p=2.18$). The fractal dimension of the
clusters, as well as the other exponents that characterize the moments
of the cluster size distribution in the vicinity of the line, are also
in good agreement with those of random percolation theory
\citep{Campi01a,Sat00}. Yet, this result was found on a deterministic
dynamical framework, without any explicit reference to a random (site
or bond) percolation mechanism. This is even more striking when the
global definition is used instead of the bond activation criterion.

The percolation line starts at the critical point, within error bars
due to the finite size of the system and to critical slowing
down. Just at this point, for the same technical reasons, geometric
critical exponents were not evaluated with enough accuracy to
distinguish between the universality classes of the Ising model and of
the random percolation. Furthermore, the total energy of the system
remains almost constant along the percolation line\footnote{The same
observation has been made about the Kert\'esz line in the lattice-gas
model \citep{Campi99}.} \citep{Campi01a}. The origin of this energy
invariance, which results from a subtle balance between internal,
center of mass kinetic, and inter-cluster potential energies (see
Eq. (\ref{parti})), is not understood \citep{Campi01b}.

It is interesting to notice that along the condensation curve, and in
particular at the critical point, self-bound clusters seem to behave
like Fisher droplets. Hence, their cluster size distributions can be
fitted by Eq. (\ref{fns}), which was proposed by Fisher. However,
self-bound clusters do not form a perfect gas as Fisher droplets do,
and there is absolutely no reason for writing their free energy as
$-\beta F_s=\ln n_s-s\ln z$. Indeed, the contribution of the
interaction between self-bound clusters to the total potential energy
of the system is at least $40\%$ along both the percolation line and
the condensation curve \citep{Sat00}. Even at very low density, such
as $\rho\simeq 0.05$ in two dimensions, the interaction between a
single monomer and a self-bound cluster markedly modifies the cluster
size distribution of the ideal gas of clusters \citep{Soto97}. In
addition, we recall that the definition of the self-bound clusters
tends to minimize interaction between clusters. These results show
strikingly that whatever microscopic definition of clusters is chosen,
it is not realistic to treat them as a perfect gas of clusters, even
at low densities $\sim \rho = \rho_c/3 \simeq 0.1$.

For a given cluster size $s$, the degrees of freedom associated with
the motion of the particles inside the clusters, and the center of
mass motion of the clusters, are respectively
characterized\footnote{These effectives temperatures, $T^{*}(s)$ and
$T^{cm}(s)$, are defined as twice the corresponding average kinetic
energy divided by the number of degrees of freedom, that is $3(s-1)$
and $3$ respectively in three dimensions.}  by an ``internal effective
temperature'' $T^{*}(s)$ and a ``translational effective temperature''
$T^{cm}(s)$. Analytical calculations \citep{Soto98a,Soto98b} as well
as numerical simulations \citep{Campi01b,Campi02} show that whatever
the size of the clusters and the thermodynamical state of the system,
$T^{*}(s)<T<T^{cm}(s)$ where $T$ is the real thermodynamic temperature
of the system. Simply put, the system can be seen as a ``hot'' gas of
``cold'' physical clusters. This result, due to the self-bound nature
of the clusters, is particularly relevant when studying the
fragmentation and expansion of a small system. Indeed, once the
clusters are isolated from each others, the internal effective
temperature becomes the real temperature of a fragment in the
microcanonical ensemble \citep{Campi02}.

As we have seen in section \ref{hill}, self-bound clusters can also be
defined by a simplified probabilistic criterion. By means of molecular
dynamics simulations of a Lennard-Jones fluid, Pugnaloni and his
collaborators \citeyearpar{Pugna02} have recently shown that this
definition strongly overestimates the percolation density by a factor
1.5-2, especially at high temperatures. In other words, probabilistic
clusters are much smaller on average than the deterministic Hill
clusters. On the other hand, the mean cluster size of the
probabilistic Hill clusters can be calculated using an integral
equation theory\footnote{Xu and Stell \citeyearpar{Xu88} used this
approach to study the percolation behaviour of a Yukawa
fluid. However, clusters are defined either by a cutoff distance or by
a probability of connection which depends on an adjustable parameter.}
\citep{Coni77a}. The percolation density is then determined by the
divergence of the mean cluster size. When compared to the molecular
dynamics results, the percolation line of the probabilistic clusters
is shifted to much lower densities and may seem to roughly coincide
with the percolation line of the deterministic clusters
\citep{Coni77a,Pugna02}.

\begin{figure}[h!]
\begin{center}
\includegraphics[angle=0,scale=.8]{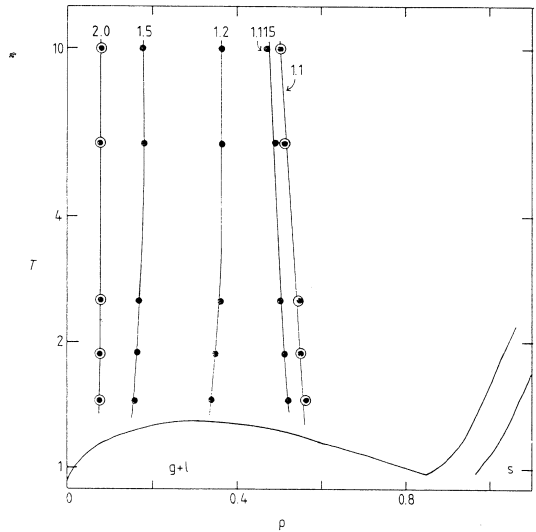}
\end{center}
\caption{\it Percolation lines of the configurational clusters in the
$T-\rho$ phase diagram of a Lennard-Jones fluid determined by
molecular dynamics simulations for two system sizes: $N=108$ (black
points) and $N=500$ (circled points). The corresponding value of the
cutoff distance $b$ is indicated for each percolation line. Note that
$b_{min} \simeq 1.12$ is the distance corresponding to the minimum of
the Lennard-Jones potential (from \citealp{Heyes88}).}
\label{fig10}
\end{figure}

For completeness, we mention that configurational clusters (see
section \ref{sec:lattice}), defined by a cutoff distance $b$, also
generate a percolation line which almost follows an isochore
($\rho=cte$), as do Ising clusters in the lattice-gas model
\citep{Geiger82,Bug85,Safran85,Xu88}. Of course, the position of the
line in the phase diagram depends crucially on the choice of $b$
\citep{Heyes88,Heyes89,Xu88}, as can be seen in figure \ref{fig10}.

\begin{figure}[h!]
\begin{center}
\includegraphics*[angle=0,scale=1.3]{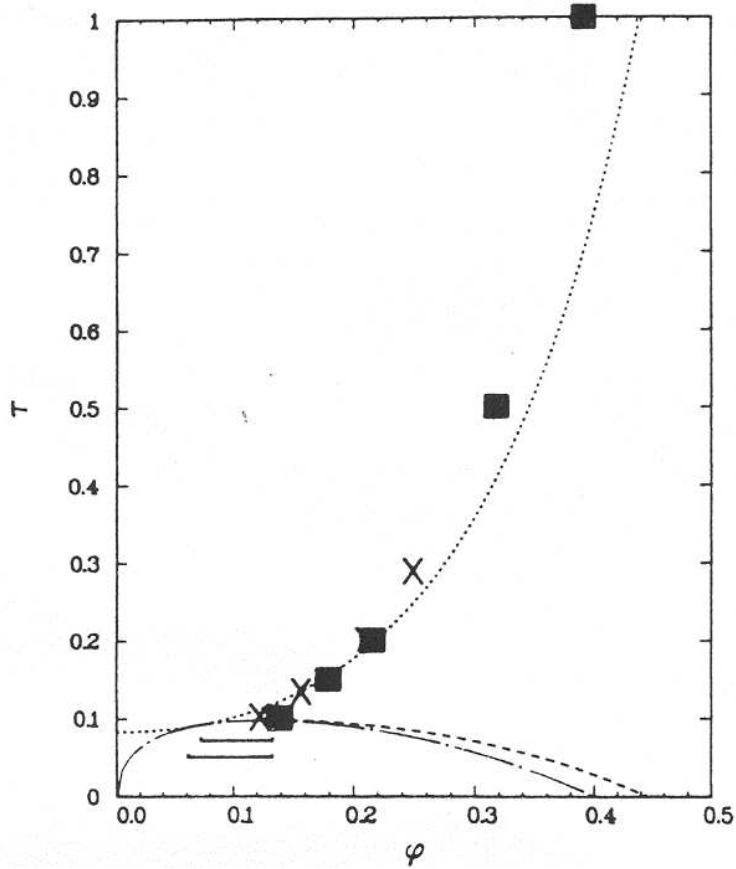}
\end{center}
\caption{\it Phase diagram of the sticky-hard-sphere model. The
stickiness parameter $\tau$ (vertical axis) is a measure of the
temperature and $\phi$ is the volume fraction (horizontal axis). The
dashed line represents the spinodal line in the Percus-Yevick
approximation. The percolation line determined by the Percus-Yevick
approximation, dotted line \citep{Chiew83}, and by numerical
simulations, black squares \citep{Kranendonk88}, and crosses
\citep{Seaton87a,Seaton87b}.  The horizontal bars give the positions
of percolation thresholds below the spinodal line (from
\citealp{Kranendonk88}).}
\label{fig11}
\end{figure}

In the very particular case of Baxter's sticky-hard-sphere model (or
adhesive-hard-sphere model) \citep{Bax68}, configurational clusters
become identical to self-bound clusters. The potential is a hard
spheres plus a square-well potential, in the limit where the depth of
the well goes to infinity, and its width goes to zero, in such a way
that the product of depth and width remains constant. The cutoff
distance $b$ is then obviously the diameter of the core. Besides, once
in contact, two particles are permanently bound. Baxter
\citeyearpar{Bax68} solved this model analytically, using the
Ornstein-Zernike equation in the Percus-Yevick approximation, and
showed that the system undergoes a liquid-gas phase transition (see
Fig. \ref{fig11}). On the other hand, Chiew and Glandt
\citeyearpar{Chiew83} used the approach proposed by Coniglio and his
collaborators \citeyearpar{Coni77a} (see section \ref{hill}) to locate
the percolation line in the phase diagram of the sticky-hard-sphere
model. As is shown in Fig. \ref{fig11}, this line does not pass by the
critical point. However, following Kranendonk and Frenkel
\citeyearpar{Kranendonk88}, we note that its location for extreme
values of the volume fraction\footnote{The volume fraction is related
to the density by $\phi=\pi \rho \sigma^3 / 6$, where $\sigma$ is the
diameter of the particles.} casts some doubt upon the adequacy of the
Percus-Yevick approximation. Indeed, the percolation line starts at
$\phi=0$, predicting the existence of a percolating cluster in vacuum,
and goes to the unphysical value $\phi=1$, greater than the close
packing volume fraction. In contrast, Monte-Carlo simulation
predictions show that the percolation line {\it starts} very close to
the critical point and goes through the supercritical phase as in the
case of a Lennard-Jones fluid
\citep{Seaton87a,Kranendonk88}. According to \cite{Kranendonk88}, the
prolongation of the percolation line under the coexistence curve
should be taken with caution.

\section{Summary and outlook}
\label{sec:conc}

Ever since Mayer proposed his theory of condensation in 1937, the
correspondence between thermodynamics and the morphology of simple
fluids in terms of clusters has attracted a lot of interest. In this
last section, we summarize the various stages of this research area,
and close by mentioning some open questions.

Initially, the purpose was to introduce clusters as tools in order to
calculate the thermodynamic quantities of a classical fluid, and in
particular to study the liquid-gas phase transition and the critical
point. To this end, condensation is defined as the formation of a
macroscopic cluster. By considering a real fluid as a gas of
non-interacting clusters, one can write the equation of state of the
system as a function of the cluster size distribution. Consequently,
the mean cluster size varies like the compressibility and then should
diverge at, and only at, the thermodynamic critical point. This is the
case by construction for Fisher droplets which provide a morphological
description of the critical point. But these phenomenological
non-interacting clusters cannot be identified as sets of particles in
the fluid. The next step was to find a microscopic definition of
clusters that behave like Fisher droplets.

In the lattice-gas model, the problem amounts to finding a definition
of clusters that produces a percolation threshold at the critical
point. Ising clusters, defined by the simple criterion of proximity in
real space, do not fulfill this condition. In contrast, Coniglio-Klein
clusters, whose definition is based on the mathematical work of
Fortuin and Kasteleyn, are explicitly designed to this end. But this
is not the end of the story: Coniglio-Klein clusters exhibit a
percolation behaviour not only at the critical point, but also along
the so-called ``Kert\'esz line'' which goes through the supercritical
region of the phase diagram. Within the context of the perfect gas of
clusters model, this percolation line should imply forbidden
thermodynamic singularities. However, because Coniglio-Klein clusters
do not form a perfect gas of clusters, there is no valid reason for
eliminating the Kert\'esz line. It remains to give a physical
interpretation of this percolation line. This is what Campi and
Krivine have done by showing that Coniglio-Klein clusters are
self-bound on average. This interpretation in terms of self-bound
clusters opens up a new field of investigation.

Self-bound clusters can be studied in the more realistic framework of
a Lennard-Jones fluid. It is interesting to notice that these clusters
were introduced not to provide a correspondence between thermodynamics
and geometry, but rather to describe processes like the fragmentation
of a piece of matter. Nevertheless, self-bound clusters strikingly
present the morphological manifestations associated with the
liquid-gas phase transition, and therefore can be considered as the
microscopic physical counterparts of the phenomenological Fisher
droplets. What is more, self-bound clusters give rise to a percolation
line in the supercritical phase. An important open question, which
arises from their definition based on physical grounds, is whether
this line could be observed in real fluids.

Various real systems behaving like simple fluids can be used to search
for signals of the self-bound clusters and the percolation line. For
example, the effusion of a fluid through a pin hole allows one to
infer the cluster size distribution from the mass yield of escaped
clusters with a given velocity. Hence, the presence of clusters
(dimers and trimers) was found in the gas phase at very low density
\citep{Miller55,Miller56}. However, this kind of experiment seems to
be inappropriate to explore a fluid at high density, in the region of
the percolation line. On the other hand, experiments dealing with
binary fluids (which belong to the universality class of the $3d$
Ising model \citep{Chen83}) were performed to give a morphological
description of the critical point in terms of clusters
\citep{Guen89}. The relationship between these clusters, defined from
persistent density fluctuations and the self-bound clusters is not yet
understood. Very recently, a sharp decrease of ${\rm H}^+$ ion
mobility in a mixture of isobutyric acid and water has been ascribed
to a percolation line of ``dynamic clusters'' which starts at the
critical point \citep{Bonn02}. At first glance, the vertical position
of this curve in the phase diagram would suggest a closer connection
with configurational clusters (see Fig. 5 in \citep{Bonn02}). However,
the slope of the percolation line of the self-bound clusters may
depend on the system considered. The question of whether these
``dynamic clusters'' are self-bound, still awaits an answer.

From a rather different perspective, the relation between clustering
and thermodynamics is critical to the determination of the phase
diagram of small systems, like atomic nuclei and aggregates
\citep{Hill63}. In a first approximation, the nuclear interaction can
be seen as a two body potential with a repulsive hard-core and a
short-ranged attraction. By disregarding quantum effects at high
enough energy, hot nuclear matter is thought to behave like a
classical simple fluid (\citealp{Bondorf95,Dasgupta01}, and references
therein). A crucial question which then arises is whether traces of a
liquid-gas phase transition can be observed in an atomic nucleus
\citep{Siemens83} ? A direct observation of such a signal is of course
quite arduous in a small system, and today there is still no
conclusive answer to this question.

To explore the phase diagram of nuclear matter, atomic nuclei are
heated and compressed by collisions at high bombarding energy (about
100 MeV/nucleon) with other nuclei. As a result, an excited nucleus
expands and breaks into fragments which can be detected using $4\pi$
multidetectors (see references in \citealp{Bondorf95}). Basically,
only the fragment size and the kinetic energy distributions can be
directly measured. The challenge is then to infer the thermodynamical
state of the system before fragmentation occurs from these two
distributions which characterize the asymptotic fragments.

To this end, several theoretical approaches have been proposed.
``Statistical Equilibrium Models'', like the SMM \citep{Bondorf95} and
the MMMC \citep{Gross97} models, assume that at a given stage of the
expansion the system is an ensemble of non-interacting spherical
fragments. Using the perfect gas of clusters model, and some elements
of nuclear physics, these models have been successful in describing
the observed fragment size distributions \citep{Botvina95}. However,
the kinetic energies are too low compared to experimental data, and
worst of all, the strongly restrictive hypotheses of the perfect gas
of clusters prevent exploration of the high density region of the
phase diagram. On the other hand, classical molecular dynamics
simulations of the expansion and fragmentation of a small system allow
one to investigate the whole phase diagram
\citep{Schlagel87,Belkacem95,Cherno01,Campi02}. In this way, it has
been shown that self-bound clusters in the hot and dense fluid are the
precursors of the observed fragments
\citep{Dorso93,Campi01b,Campi02}. In particular, self-bound clusters
in the initial system and asymptotic fragments have the same size
distribution. The initial thermodynamic state could be deduced from
the measured quantities by using the correspondence between the phase
diagram and the self-bound cluster size distribution at equilibrium
\citep{Campi02}. It must be emphasized that various experiments of
fragmentation of atomic nuclei\footnote{See for example
\citep{Finn82,Campi88,Gilkes94,Schu96,Campi00,Bauer02}.} and atomic
aggregates \citep{Fari98,Gobet01} exhibit a percolation behaviour of
the fragment size distribution (U-shapes, power law, and exponentials
at respectively low, intermediate, and high energy).

Finally, beyond the framework of simple fluids, attractive colloidal
suspensions display aggregation and dynamic behaviours that could be
understood in terms of self-bound clusters. Compared with atoms,
colloidal particles have larger sizes, a slower diffusion, and a range
of attraction much smaller than the hard-core diameter
\citep{Pusey91,Anderson02}. These features contribute to form
morphological structures that can be visualized by using microscopic
imaging \citep{Segre01}. We are not aware of a microscopic theoretical
description of these clusters. What is more, percolation lines are
frequently observed in attractive colloidal suspensions. According to
the particular physical situation, the percolation threshold is
associated with a sharp change in the following quantities
\citep{Coni01}: i) Electrical conductivity in water-in-oil
microemulsions \citep{Chen94,Weigert97}, ii) Viscosity of triblock
copolymer micellar solutions \citep{Mallamace99,Mallamace01}, iii)
Elastic modulus of dispersions of silica spheres grafted with polymer
chains \citep{Grant93}, and iv) Correlation functions, measured by
dynamic light scattering, of dispersions of silica spheres coated with
stearyl alcohol \citep{Verd95}. It is worth noting that in all these
cases the percolation line passes in the vicinity of the critical
point. Therefore, clustering seems to be intimately related to the
dynamic properties of complex fluids. Clearly, more work is needed in
this direction.

\section*{Acknowledgments}
I am deeply grateful to my collaborators Xavier Campi and Hubert
Krivine. During the writing of this article, I have benefited from
fruitful discussions with Antonio Coniglio to whom I am thankful. I
wish also to thank Donald Sprung for making useful remarks about the
manuscript. This work was supported by grants from NSERC Canada
(research grant SAPIN-3198) and from the European TMR Network-Fractals
(Postdoctoral Grant, Contract No. FMRXCT-980183).

\appendix

\section{Perfect gas of clusters model}
\label{sec:pgc}

Let us consider a set of non-interacting clusters that do not have any
volume. The $n_{s}$ indistinguishable clusters of size $s$ form a
chemical species characterized by their mass $ms$, their chemical
potential $\mu_{s}$ and the partition function $q_{s}(T,V)$.

In the canonical ensemble, at constant volume and temperature, the
differential of the free energy is 
$$
dF=\sum_{s} \mu_s dn_s.
$$ 
Furthermore, mass conservation is ensured by the condition $\sum
sn_s=N$, or $\sum sdn_s=0$. Chemical equilibrium between clusters
implies the minimization of the free energy. By the method of
Lagrangian multipliers, we obtain
\begin{equation}
\label{e22} \mu_{s}=s\mu \qquad \textrm{for} \qquad s=1,2,\dots,N 
\end{equation}
where $\mu$ is the particle chemical potential. The fugacity $z_s$ of
clusters of size $s$ is then
$$ 
z_s=e^{\beta \mu_s}=e^{\beta s \mu }=z^{s}
$$
where $z=z_1$ is the fugacity of the particles. The partition function
for a given partition $\vec P=\{n_1,n_2,... \}$ of particles into
non-interacting clusters is given by
\begin{equation}
\label{qfb} Q_{\vec P}(T,V)=\prod_{s=1}^{\infty}
\frac{q_{s}^{n_{s}}}{n_{s}!}.
\end{equation}
The total partition function of a system of $N$ particles is then
$$
 Q_{N}(T,V)=\sum_{\vec P}Q_{\vec P}\delta(\sum sn_s-N)
$$
where $\sum_{\vec P}$ is a sum over all the possible partitions,
whatever the number of particles. By moving over to the grand
canonical ensemble, the grand partition function $\Xi(z,T,V)$ can be
written as a function of the fugacity $z$:
\begin{eqnarray*}
\Xi(z,T,V) &=& \sum_{N \ge 0}z^N Q_N \\
&=& \sum_{N \ge 0}z^N \sum_{\vec P} Q_{\vec P} \delta (\sum sn_s -N) \\
&=& \sum_{N \ge 0} \sum_{\vec P} z^N \prod_{s=1}^{\infty} 
\frac{{q_s}^{n_s}}{n_{s}!} \delta (\sum sn_s -N).\\ 
\end{eqnarray*}
By summing over $N$, we obtain
\begin{eqnarray}
\Xi(z,T,V) &=& \sum_{\vec P} \prod_{s=1}^{\infty} z^{sn_{s}}
\prod_{s=1}^{\infty} \frac{q_{s}^{n_{s}}}{n_{s}!} \nonumber\\ &=&
\sum_{\vec P} \prod_{s=1}^{\infty} \frac{(q_{s}z^s)^{n_{s}}}{n_{s}!}
\nonumber \\ &=& \prod_{s=1}^{\infty}
\sum_{n_s=0}^{\infty}\frac{(q_{s}z^s)^{n_{s}}}{n_{s}!} \nonumber\\ &=&
\prod_{s=1}^{\infty} e^{q_{s}z^{s}} \nonumber\\ &=&
e^{\sum_{s=1}^{\infty}q_{s}z^{s}}. \label{eparti}
\end{eqnarray}
The grand partition function $\Xi(z,T,V)$ allows us to calculate the
grand potential $\Omega=F-\mu N=-kT\ln \Xi(z,T,V)$.  Because
$\Omega=-PV$ (see for example \citealp{Landau59}), we infer the pressure
and the density in the thermodynamic limit:
\begin{eqnarray}
\beta P & \equiv & \lim_{V \rightarrow \infty} (\frac{1}{V} \ln
\Xi) = \lim_{V \rightarrow \infty}( \frac{1}{V}\sum_{s=1}^{\infty}
q_{s}(T,V) z^{s} ) \label{fbpkt} \\ \rho & \equiv & \lim_{V
\rightarrow \infty} (\frac{z}{V}\frac{\partial \ln \Xi}{\partial z}) =
\lim_{V \rightarrow \infty}( \frac{1}{V}\sum_{s=1}^{\infty}
sq_{s}(T,V) z^{s}). \label{fbro}
\end{eqnarray}
For large enough volume, ${q_{s}(T,V)}/{V}$ becomes independent of $V$
and tends to $\widetilde{q}_{s}(T)$. In the domain of convergence of
series (\ref{fbpkt}) and (\ref{fbro}), we can permute the limit when
$V \rightarrow \infty$ and the sum over $s$. The pressure and density
expression are then given by
\begin{eqnarray*}
\beta P &=& \sum_{s=1}^{\infty} \widetilde{q}_{s}(T) z^{s}\\ \rho & =
& \sum_{s=1}^{\infty} s\widetilde{q}_{s}(T) z^{s}.
\end{eqnarray*}
It is convenient to express the thermodynamic quantities as functions
of the series $\pi(T,z)$ \citep{Fish67a}:
\begin{eqnarray*}
\pi(T,z)& \equiv &\sum_{s=1}^{\infty} \widetilde{q}_{s}(T) z^{s} \\
\pi^{(n)}(T,z)& \equiv &(z\frac{\partial}{\partial z})^{(n)}\pi(T,z).
\end{eqnarray*}
The expressions for pressure, density, compressibility, and specific
heat at constant volume, become
\begin{eqnarray*}
\beta P &=& \pi(T,z) \\ \rho &=& \pi^{(1)}(T,z) \\
\chi_{T}&\equiv&\frac{1}{\rho}\frac{\partial \rho}{\partial
P}=\frac{1}{\rho}\frac{\partial \rho}{\partial z} \frac{\partial
z}{\partial P}=\frac{\beta}{\rho^{2}}\pi^{(2)}(T,z) \\ C_{v} &=& T
{\frac{\partial S}{\partial T}}_{V} = -T {\frac{\partial^{2}
\Omega}{\partial T^{2}}}_{V,\mu}=kT^{2}V{\frac{\partial^{2}
\pi(T,z)}{\partial T^{2}}}_{V,\mu}.
\end{eqnarray*}
Let us calculate the cluster size distribution. From now on, $n_s$
is the {\it mean} number of clusters of size $s$:
$$
n_{s}=z_{s}\frac{\partial \ln \Xi}{\partial z_{s}}. 
$$
From Eq. (\ref{e22}) and Eq. (\ref{eparti}) we obtain
$$ 
n_{s}=z_s q_{s}=q_{s}z^{s}. 
$$
Define $m_k$ as the moment of order $k$ of the cluster size
distribution:
$$
m_k=\sum_{s=1}^{\infty} s^k n_s .
$$ 
In this theoretical framework, thermodynamic quantities are associated
with the moments $m_k$. Indeed,
\begin{eqnarray*}
\pi(T,z)^{(n)} &=& \frac{1}{V} \sum_{s=1}^{\infty} s^{n} q_{s}z^{s} \\ 
&=& \frac{1}{V} \sum_{s=1}^{\infty} s^{n}n_{s}=\frac{m_{n}(T,z)}{V}.
\end{eqnarray*}
We then have
\begin{eqnarray} 
\beta P &=& \frac{m_{0}(T,z)}{V} \label{moka1} \\
\rho &=& \frac{m_{1}(T,z)}{V} =\frac{N}{V}\label{moka2}  \\  
\chi &=& \frac{\beta m_{2}(T,z)}{V\rho^{2}} =
\frac{V}{kT}\frac{m_{2}(T,z)}{m^{2}_{1}(T,z)} \label{moka3} \\
C_{v} &=& kT^{2}{\frac{\partial^{2} m_{0}}{\partial
T^{2}}}_{V,\mu}. \label{moka4}
\end{eqnarray}
From Eqs. (\ref{moka1}) and (\ref{moka3}) it follows that\footnote{In
a perfect gas of {\it particles}, there are only monomers:
$m_{0}=m_{1}=m_{2}=N$ and $\chi=1/P$.}:
$$
P\chi=\frac{m_{0}m_{2}}{m_{1}^{2}}=\gamma_{2}
$$
where $\gamma_{2}$ is a quantity that diverges as $m_2$ at the
percolation threshold \citep{Campi88}.

Equations (\ref{moka1})-(\ref{moka4}) clearly show that in a perfect
gas of clusters model, geometric and thermodynamic quantities must
diverge at the same points of the phase diagram. In particular, the
second moment of the cluster size distribution diverges like the
compressibility.
 
\section{Stability of the probabilistic Hill clusters}
\label{stabhill}

To begin, we define the stability against particle emission. The
separation energy $e_i^{C_s}$ is the energy of a particle $i$
calculated in the center of mass system of the cluster $C_s$ to which
it belongs. A cluster is stable against particle emission if, for each
particle $i$ belonging to $C_s$, we have:
$$
e_{i}^{C_s} \equiv K^{*}_{i}+ \sum_{j \in C_s, j \ne i}u({r_{ij}}) < 0
$$ 
where $K^{*}_{i}$ is the kinetic energy of particle $i$ calculated in
the center of mass system of cluster $C_s$, that is
$$
K^{*}_{i}=\frac{m}{2}(\vec v_{i} -\vec V^{C_s})^2,
$$
and $\vec V^{C_s}=(\sum_{i \in C_s} \vec v_{i})/s$ is the velocity of
the center of mass of the cluster $C_s$. When the positions and
velocities of the particles belonging to a given cluster are known,
this criterion is straightforward to check.

\begin{figure}[h!]
\begin{center}
\includegraphics[angle=0,scale=.5]{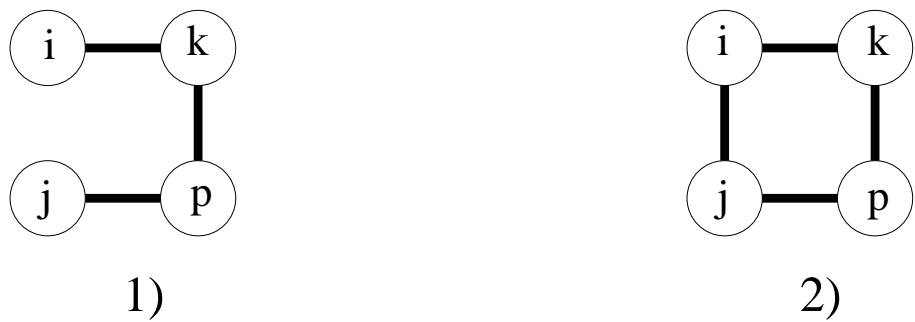}
\end{center}
\caption{\it Particle $i$ interacts with two particles of the cluster
it belongs to.  $1)$ Particle $i$ is linked to the cluster by only one
Hill's bond (heavy line) with particle $k$, but interact also with
particle $j$. $2)$ Particle $i$ is linked to the cluster by two Hill's
bonds to particles $k$ and $j$.}
\label{fig12}
\end{figure}

Let us now show that probabilistic Hill clusters are stable on average
against particle emission in the lattice-gas model \citep{Campi97}. To
this end, we evaluate the average sign of the separation energy. For
large enough clusters, the velocity of the center of mass can be
neglected. The relative kinetic energy of a particle is then equal to
$mv_i^2/2$, where $v_i$ is randomly given by the Maxwell distribution
of variance $1/\beta m$. First consider the case of a particle
interacting with only one other particle of the cluster. Its
separation energy is then given by
$$
e_i=\frac{m}{2}v_i^2-\epsilon.
$$
Because the particle belongs to the cluster, we have by definition of
a Hill's bond $e_i<0$. Now, suppose the particle interacts with two
particles of the cluster (see Fig. \ref{fig12}). The separation energy
of the particle is
$$
e_i=\frac{m}{2}v_i^2-2\epsilon.
$$
The probability that $e_i$ is negative is $Pr(mv_i^2/2 \le
2\epsilon)=p_{Hill}(2\beta \epsilon)$. On the other hand, the
probability that the particle is linked to the cluster either by one
or two independent bonds is given by
$$
2p_{Hill}(\beta\epsilon)-p_{Hill}^2(\beta\epsilon).
$$
According to Eq. (\ref{hilck}), we can substitute $p_{ck}$ for
$p_{Hill}$. It is then very simple to check that
$$
2p_{ck}(\beta\epsilon)-p_{ck}^2(\beta\epsilon)=p_{ck}(2\beta \epsilon).
$$ 

The generalization to more bonds between the particle and the cluster
is straightforward. In conclusion, if a particle belongs to a
probabilistic Hill cluster, its separation energy is negative. The
condition for stability against particle emission is fulfilled on
average.

\end{document}